\documentclass[numberedappendix]{emulateapj}

\usepackage{apjfonts}
\usepackage{lscape}

\shortauthors{Mao et al.}
\begin{document}
\title{A Radio and Optical Polarization Study of the Magnetic Field in the Small Magellanic Cloud}
\submitted{Accepted to {\em The Astrophysical Journal}}
\author{S. A. Mao,\altaffilmark{1} 
B. M. Gaensler,\altaffilmark{2,8}
S. Stanimirovi\'{c},\altaffilmark{3}
M. Haverkorn,\altaffilmark{4,9}
N. M. McClure-Griffiths,\altaffilmark{5}
L.~Staveley-Smith\altaffilmark{6,10}
and J. M. Dickey\altaffilmark{7}}

\altaffiltext{1}{Harvard-Smithsonian Center for Astrophysics, Cambridge, MA 02138; samao@cfa.harvard.edu}
\altaffiltext{2}{School of Physics, The University of Sydney, NSW 2006, Australia}
\altaffiltext{3}{Department of Astronomy, University of Wisconsin, Madison, WI 53706}
\altaffiltext{4}{Astronomy Department, University of California, Berkeley, CA 94720}
\altaffiltext{5}{Australia Telescope National Facility, CSIRO, Epping, NSW 1710, Australia}
\altaffiltext{6}{School of Physics, University of Western Australia, Crawley, WA 6009, Australia}
\altaffiltext{7}{Physics Department, University of Tasmania, Hobart, TAS 7001, Australia}
\altaffiltext{8}{Alfred P. Sloan Research Fellow, Australian Research Council Federation Fellow}
\altaffiltext{9}{Jansky Fellow, National Radio Astronomy Observatory}
\altaffiltext{10}{Premier's Fellow}

\clearpage

\begin{abstract}
We present a study of the magnetic field of the Small Magellanic
Cloud (SMC), carried out using radio Faraday rotation and optical
starlight polarization data.  Consistent negative rotation measures
(RMs) across the SMC indicate that the line-of-sight magnetic field
is directed uniformly away from us with a strength $0.19 \pm
0.06$~$\mu$G. Applying the Chandrasekhar-Fermi method to
starlight polarization data yields an ordered magnetic field in the
plane of the sky of strength $1.6 \pm 0.4$~$\mu$G oriented at a
position angle $4^\circ \pm 12^\circ$, measured counter-clockwise
from the great circle on the sky joining the SMC to the Large
Magellanic Cloud (LMC).  We construct a three-dimensional magnetic
field model of the SMC, under the assumption that the RMs and
starlight polarization probe the same underlying large-scale field.
The vector defining the overall orientation of the SMC magnetic
field shows a potential
alignment with the vector joining the center of the SMC to the center of the LMC, suggesting the possibility of a ``pan-Magellanic''
magnetic field. A cosmic-ray driven dynamo is the most viable
explanation of the observed field geometry, but has difficulties
accounting for the observed uni-directional field lines. A study
of Faraday rotation through the Magellanic Bridge is needed to
further test the pan-Magellanic field hypothesis.

 \end{abstract}


\section{Introduction}
\label{section:introduction}
Magnetic fields play key roles in many astrophysical processes in the interstellar medium (ISM) --- they accelerate and confine cosmic rays, trigger star formation and exert pressure to balance against gravity \citep{beck2007}. Therefore, to better understand galaxy evolution, investigating the structure, origin and evolution of galactic magnetic fields is necessary.

It is useful to picture the total magnetic field at any location in a galaxy as a superposition of an ordered large-scale component and a random small-scale component. An ordered (or uniform) field can either be coherent or incoherent: a coherent field has unidirectional field lines, whereas an incoherent field has field lines of the same orientation but has frequent field reversals. A large-scale dynamo is the only known mechanism that can generate large-scale coherent fields \citep{beck2004a}.

Coherent magnetic fields have been observed in normal spiral galaxies such as the Milky Way and M31.
The fields are typically in spiral-like configurations with field strengths of a few $\mu$G \citep{beck2007}. Because these galaxies have significant differential rotation, such observations can be explained by the standard $\alpha$-$\omega$ dynamo which amplifies and orders the field by small scale turbulent motion (the $\alpha$-effect) as well as differential rotation (the $\omega$-effect) in the galactic disk on a global e-folding time of  $\sim$ 10$^9$ yrs \citep{shukurov2004}. Despite its success in accounting for the large scale coherent field seen in spiral galaxies, the standard dynamo theory fails to explain the presence of coherent magnetic fields discovered in several irregular galaxies such as NGC 4449 and the Large Magellanic Cloud (LMC), due to its long amplification time scale \citep{klein1996,chyzy2000,gaensler2005}. 

The polarized radio continuum emission of NGC 4449, a dwarf irregular galaxy, at 4.9 and 8.6 GHz reveals large scale spiral like structure in the magnetic field. Moreover, this slowly rotating galaxy shows regions of coherent magnetic field from Faraday rotation studies \citep{klein1996}. The long field amplification time scale of the classical mean field dynamo argues against this being the underlying mechanism that produces the magnetic field observed in NGC 4449. A Faraday rotation measure study of extragalactic polarized sources behind the LMC carried out by \cite{gaensler2005} suggests that the LMC hosts a coherent axisymmetric magnetic field of strength $\sim$ 1$\mu$G. The random component dominates over the ordered component with a strength of $\sim$ 3 $\mu$G. It is believed that close encounters between the Magellanic Clouds and the Milky Way have triggered episodes of star formation in the LMC over the past 4 billion years \citep{bekki2005}. Any coherent field built up by the standard dynamo would have been disrupted by the outflow from active star forming regions. Hence, the existence of a coherent field in the LMC suggests that a field generation mechanism with a much faster amplification time scale is at work.

 A much more efficient process, the 
\cite{parker1992} dynamo, could account for the large-scale magnetic fields detected in irregular galaxies such as NGC 4449 and the LMC \citep{hanasz2004}. Vertical pressure from cosmic rays can force magnetic field lines into the galactic halo and form loops which reconnect and then are amplified by the $\omega$ effect. This process can significantly increase the $\alpha$ effect and can operate over a much shorter amplification time scale than the $\alpha-\omega$ dynamo \citep{hanasz2004}.

As a close neighbor of the Milky Way, the large angular extent of the Small Magellanic Cloud (SMC) on the sky allows us to determine RMs of  polarized background radio sources whose projections lie behind it. Since RM is an integral of the line of sight magnetic field strength weighted by the thermal electron density, only a coherent field can produce a consistent sign in RM as a function of position on the sky. An RM study can distinguish between coherent and incoherent fields and can indicate the geometry of any coherent field, and therefore can potentially reveal the field generation mechanism.

The alignment of non-spherical dust grains with the magnetic field in the ISM linearly polarizes optical radiation that travels through it. Therefore, measuring the optical polarization of stars in the SMC enables us to estimate the orientation of the ordered magnetic field in the plane of the sky. The spread in polarization position angle of an ensemble of starlight polarization measurements allows one to estimate the mean strength of the ordered component of the magnetic field using the Chandrasekhar-Fermi (\citeyear{chandrasekhar1953}), or C-F method. Assuming that the field is unidirectional and knowing both the line-of-sight and the plane-of-the-sky magnetic field strength and orientation, one can construct a three dimensional magnetic field vector for the SMC, which can further constrain the field generation mechanism.

In this paper, we present the results of a radio and optical polarization study of the magnetic field in the SMC. We use RMs of polarized extragalactic background radio sources to determine the magnetic field strength and direction along the line of sight. The orientation and strength of the plane-of-the-sky magnetic field is studied using optical polarization of stars in the SMC. We start in \S~\ref{subsection:smc} by reviewing the properties of the SMC and we summarize previous studies on SMC's magnetism in \S~\ref{subsection:previousstudies}. In \S~\ref{subsection:faradayrotation} and \S~\ref{subsection:osp}, we summarize the physics behind the RM method and the C-F method respectively. We then describe the observation, data reduction procedures and present results in \S~\ref{section:obsdataresults}. We derive the line-of-sight magnetic field of the SMC in \S~\ref{section:wimbfield}. In \S~\ref{section:posb}, we estimate the plane-of-the-sky magnetic field and the random field strength in the SMC. A 3D magnetic field vector of the SMC is constructed in \S~\ref{section:3Dfield}. A discussion of possible field generation mechanisms is provided in \S~\ref{section:discussion}. 

Throughout this paper, we represent physical quantities in the plane of the sky by the subscript $\perp$ and those along the line of sight by the subscript $\parallel$. We denote the average of a quantity $x$ over the plane of the sky and that averaged along the line of sight by $\langle x \rangle$ and $\bar{x}$ respectively. Table~\ref{table:glossary} contains a glossary of the variables used in this paper.

\subsection{The Small Magellanic Cloud}
\label{subsection:smc}

The Small Magellanic Cloud is a nearby gas rich dwarf irregular galaxy. Recent precise measurements of apparent magnitudes of stars at the tip of the red giant branch in the SMC yield a distance modulus of 18.99$\pm$ 0.03(formal)$\pm$0.08(systematic)  \citep{cioni2000}, which corresponds to a distance of roughly 63 $\pm$ 1 kpc. In this paper, we adopt a distance to the SMC of 60 kpc. Basic parameters of the SMC are listed in Table~\ref{table:smcproperties}. Both the SMC and the LMC are thought to be satellite galaxies of the Milky Way. However, a recent study by \cite{besla2007} suggests that the Clouds are not bound to the Milky Way but are on their first passage about the Galaxy. It is still of great debate as to whether the Magellanic Clouds formed as a binary, or whether they became dynamically coupled to each other $\sim$ 4 Gyrs ago \citep{bekki2005}. The most recent proper motion measurements of the Clouds suggest that both scenarios are equally probable \citep{kallivayalil2006,piatek2007}. It is believed that the last close encounter of the Magellanic Clouds $\sim$ 0.2 Gyrs ago triggered star formation in the SMC and created the morphological and kinematic features seen in the present day SMC \citep{yoshizawa2003}.

 \cite{stanimirovic2004} found that the depth of the SMC is within its tidal radius ($\sim$ 4 $-$ 9 kpc). \cite{lah2005} have measured distances to pulsating red giants in the SMC and found a distance scatter of 3.2 $\pm$ 1.6 kpc, which agrees with the results of \cite{stanimirovic2004}. N-body simulations of the gravitational interaction between the LMC, SMC and the Milky Way have been able to reproduce the large line-of-sight extent of the SMC and its two tidal arms \citep{gardiner1994,yoshizawa2003}. 

The gas component in the SMC shows signs of rotation whereas the old stellar component does not \citep{hatzidimitriou1997}. The gas kinematics of the SMC were investigated by \cite{stanimirovic2004}, who found a strong velocity gradient in HI across the SMC from the southwest to the northeast. \cite{stanimirovic2004} constructed a rotation curve of the gas disk, and derived a maximum rotation velocity of  50 km s$^{-1}$.

\subsection{Previous Studies of Magnetism in the SMC}
\label{subsection:previousstudies}
The most common way to study magnetic fields in external galaxies is by observing synchrotron emission at radio wavelengths. \cite{haynes1986} examined linear polarization maps of the SMC at 1.4 GHz. They found, without any Faraday rotation correction, an ordered magnetic field directed along the SMC's bar in the plane of the sky. \cite{loiseau1987} analyzed radio continuum maps of the SMC at  408 MHz, 1.4 GHz and 2.3 GHz and obtained a total equipartition field strength of $\sim$ 5 $\mu$G by using an average non-thermal spectral index $\alpha$ of 0.87 (specific intensity of synchrotron emission  $I_\nu\propto\nu^{-\alpha}$ ) and a depth of the synchrotron emitting region of 6 kpc. \cite{haynes1990} observed the SMC at 2.5, 4.8 and 8.6 GHz and concluded that the SMC has a large-scale magnetic field, since weak polarized emission is detected across the whole SMC body.

 \cite{chi1993} measured the $\gamma$-ray flux from the SMC to determine the field strength from radio synchrotron emission without needing to invoke the equipartition assumption. They obtained an estimate which exceeded the equipartition value and concluded that energy equipartition is not valid in the SMC.
\cite{pohl1993}, however, took the energy density in cosmic ray electrons into account, and demonstrated that energy equipartition between magnetic field and cosmic rays is not necessarily violated. 

One should note that calculating the total equipartition field requires knowledge of the depth of the synchrotron emitting layer of the SMC and the inclination of magnetic field with respect to the plane of the sky, which are both poorly constrained in the SMC. Also, as explained by \cite{beck2005}, the classical equipartition energy formula underestimates the true equipartition field strength, since the former involves  integrating the radio spectrum with a fixed frequency interval instead of a fixed energy interval, and with insufficient knowledge of the ratio of the total energy density of cosmic ray nuclei to that of the electrons and positrons. We will further explore this issue in \S~\ref{subsection:equipartition}.

\begin{figure}
\centerline{
\includegraphics[width=0.8\textwidth]{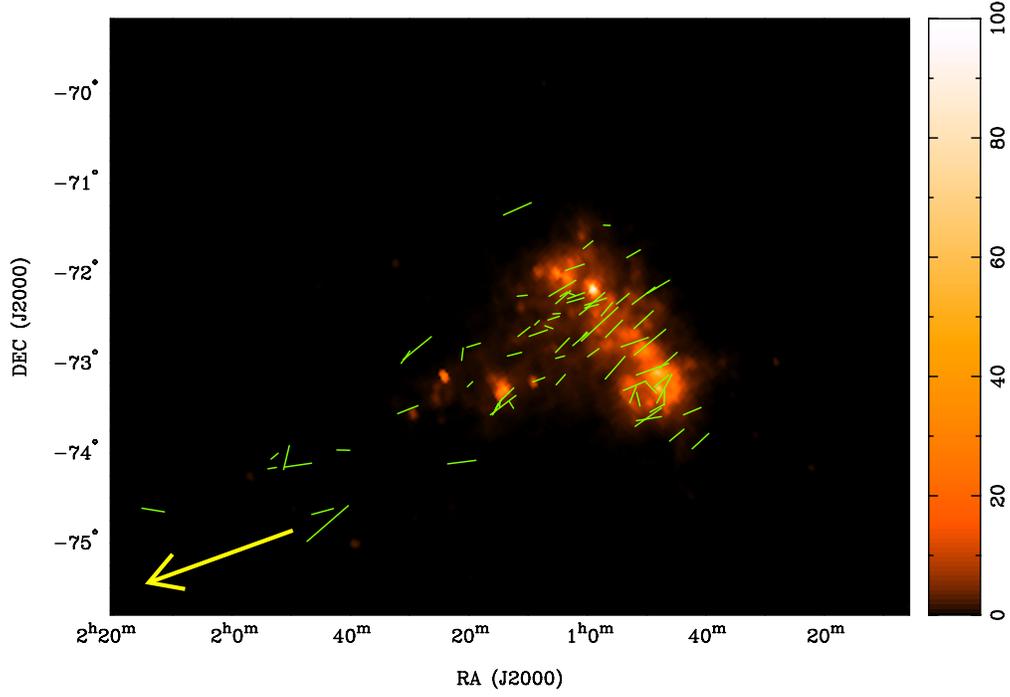}}
\caption{Optical starlight polarization ``vectors", before foreground correction, towards stars in the SMC and in part of the Magellanic Bridge \citep{mathewson1970a}, overlaid on an \emph{IRAS} 60 $\mu$m image. The color scale to the right of the image is the IR flux in units of MJy per steradian. The orientation of the line segments indicates the polarization position angle whereas the length of the line segments is proportional to the polarized fraction. The eastmost line segment corresponds to a star with observed polarized fraction of 0.51\%. Each star is located at the center of its corresponding line segment. Stars are indicated at their positions as of epoch 1975, as given in \cite{mathewson1970a}. The yellow arrow indicates the direction towards the LMC.}
\label{fig:starlight}
\end{figure}

Optical polarization from stars in the SMC can be used to map the geometry of the plane-of-the-sky component of the magnetic field, assuming that the observed polarization is due to scattering by non-spherical foreground dust grains aligned by the local magnetic field. Polarization measurements for 147 SMC stars have been made by \cite{mathewson1970a,mathewson1970b}, \cite{schmidt1970,schmidt1976}, and \cite{magalhaes1990}. Since the polarization ``vectors" \footnotemark[1] \footnotetext[1]{Position angles provide information on the orientation of the polarization plane, but not the direction. Hence, there is a 180$^\circ$ direction ambiguity.}, after removal of Galactic foreground  polarization,  appear to run parallel to the direction connecting the Magellanic Clouds (Figure~\ref{fig:starlight}), the ``Pan-Magellanic" magnetic field hypothesis emerged, which suggests the existence of a large scale magnetic field associated with the entire Magellanic system. \citeauthor{wayte1990} 's  (\citeyear{wayte1990}) reanalysis of previously obtained starlight polarization data sets appeared to support the idea of this Pan-Magellanic magnetic field. However, as the Galactic foreground polarization also runs along the projection of the line joining the Magellanic Clouds \citep{schmidt1970}, any contribution from Galactic foreground that has not been correctly subtracted could be misinterpreted as an intrinsic magnetic field connecting the Magellanic Clouds. In addition, anisotropic scattering in the ISM may also polarize starlight \citep{widrow2002}. Therefore, one has to be cautious interpreting these results. In \S~\ref{subsection:starlightcf}, we will further analyze the optical starlight polarization data using the C-F method to derive the ordered magnetic field strength of the SMC in the plane of the sky.

\subsection{Faraday Rotation }
\label{subsection:faradayrotation}
When linearly polarized light travels through a magnetized plasma, the plane of polarization rotates due to birefringence. The change in the polarization position angle $\Delta$$\phi$ in radians is given by
\begin{equation}
\Delta\phi = {\rm RM}  \lambda^{2}
\end{equation}
where $\lambda$ is the wavelength of the radiation measured in meters and RM is the rotation measure, defined by
\begin{equation}
{\rm RM} =0.812 \int ^{observer}_{source} {n_{e}(l){B_{\parallel} (l)}} dl~~~\rm{rad~m^{-2}}
\label{eq:rmdef}
\end{equation}
In the above equation, $n_e(l)$ (in cm$^{-3}$) is the thermal electron density, $B_{\parallel}(l)$ (in $\mu$G) is the line of sight magnetic field strength and d$\it{l}$ (in pc) is a line element along the line of sight. The sign of the RM gives the direction of the line of sight component of the average field. For example, a negative RM represents a field whose line of sight component is directed away from us.

RMs for extragalactic radio sources behind the SMC can be decomposed into various contributions along the line of sight: the intrinsic RM of the source, the RM through the intergalactic medium (IGM), the RM through the SMC, and the foreground Milky Way RM.
\begin{equation}
\rm{RM_{observed}}=\rm{RM_{Intrinsic}+RM_{IGM}+RM_{SMC}+RM_{Milky Way}}
\label{eq:rmsum}
\end{equation}

$\rm{RM_{Milky Way}}$ can be estimated by observing RMs of extragalactic sources whose projections on the sky lie outside, but close to the SMC. RMs of extragalactic sources at Galactic latitudes $|b| > 30^\circ$ have a standard deviation $\sim$ 10 rad m$^{-2}$  \citep{johnstonhollitt2004b}. In addition, \cite{broten1988}  showed that the extragalactic RMs in the neighborhood of the SMC have ${\rm |RM_{Intrinsic}+RM_{IGM}+RM_{Milky Way}|}$ $\le$ 25 rad m$^{-2}$. This implies that the intrinsic RM and the RM through the IGM are both small compared to the statistical errors of our RM measurements (See Table~\ref{table:rawrms}). After the removal of the Galactic foreground, the observed RM should adequately represent the RM through the SMC.  Ionospheric Faraday rotation may also contaminate our data. However, since the magnitude of RM induced by the ionosphere is typically only  $\sim$ 1 rad m$^{-2}$ \citep{tinbergen1996}, this is not of great concern in our experiment as the statistical errors of our RM measurements greatly exceed this value (See Table~\ref{table:rawrms}).

Faraday rotation is complementary to other measurement techniques such as equipartition, synchrotron intensity and starlight polarization since RMs provide the direction of the magnetic field (and hence the field coherency), while other techniques only provide the field orientation and estimates of the field strength, but not its direction. With independent knowledge of the thermal electron density and the line of sight depth of the SMC, one can estimate the average line-of-sight magnetic field strength using Equation~(\ref{eq:rmdef}) assuming that there is no correlation between electron density and magnetic field on small scales. If such correlation or anti-correlation exists, it will result in either underestimation or overestimation of the field strength by a factor of up to two to three \citep{beck2003}. 

\subsection{Optical Starlight Polarization and the Chandrasekhar-Fermi Method}
\label{subsection:osp}

Starlight polarization alone does not directly give the magnetic field strength. However, measuring the spread in polarization position angles for an ensemble of stars allows one to estimate the mean strength of the ordered component of the magnetic field \citep{chandrasekhar1953}. This technique assumes that the magnetic field is frozen into the gas and that turbulence leads to isotropic fluctuation of the magnetic field around the mean field direction \citep{heitsch2001,sandstrom2001}. Assuming equipartition between the turbulent kinetic and the magnetic energy, the ordered magnetic field strength averaged over the plane of the sky, $\langle B_{o,\perp}  \rangle$, is given by \citep{heitsch2001}: 
\begin{equation}
\label{eq:cfmethod}
\langle B_{o,\perp} \rangle ^2  = 4 \pi \rho \frac { \sigma_{v_{los}}^2} { \sigma (\rm {tan}~ \delta_p )^2 }
\end{equation} where $\rho$ is the density of the medium, $\theta_{p}$ is the measured polarization position angle, $\langle \theta_{p} \rangle$ is the weighted mean of the measured position angles, $\delta_{p} \equiv {\theta_{p}} - \langle \theta_{p} \rangle$, and $\sigma_{v_{los}}$ is the dispersion of the line-of-sight velocity in the medium. 

\section{Observations, Data Reduction and Results}
\label{section:obsdataresults}

\subsection{Radio Observations}
\label{subsection:radiodata}

RM data were acquired at the Australia Telescope Compact Array (ATCA) over the period 2004 July 10th --18th, using the 6A array configuration spanning baselines from 336.7 m to 5938.8 m, with a total of  32 adjacent frequency channels each of bandwidth 4 MHz centered on 1384 MHz. The standard primary flux calibrator PKS B1934$-$638, whose flux at 1384 MHz was assumed to be 14.94 Jy, was observed at the beginning and the end of each observation. The secondary calibrator PKS B0252-712 was observed every hour and was used to correct for polarization leakages and to calibrate the time-dependent antenna gains. To cover the whole SMC as well as the region around it, we scanned a 40-square-degree region divided into 440 pointings. For each pointing, we obtained 30 cuts of 30 seconds, resulting in a total observing time of 110 hours. These observations have poor sensitivity on scales larger than $\sim$ 30 arcseconds. Therefore, extended sources in the SMC and diffuse emission from the SMC itself are not detected; what we mainly see are background point sources.

The MIRIAD package was used for data reduction \citep{sault2003}. Data were first flagged and calibrated. Flagging and rebinning the 32 4MHz-wide channels resulted in 13 8-MHz wide channels. For each pointing and frequency channel, maps of Stokes parameters $Q$ and $U$ were made. These maps were then deconvolved using the CLEAN algorithm. A final map was generated by convolving the sky model with a Gaussian beam of dimensions 13''$\times$8" . We produced a restored image for each pointing, for Stokes $Q$ and $U$ at each of the 13 frequency channels. This results in a total of 11440 images, each with a sensitivity of $\sim$ 1 mJy per beam. 

\begin{figure}
\centering
\includegraphics[width=0.4\textwidth]{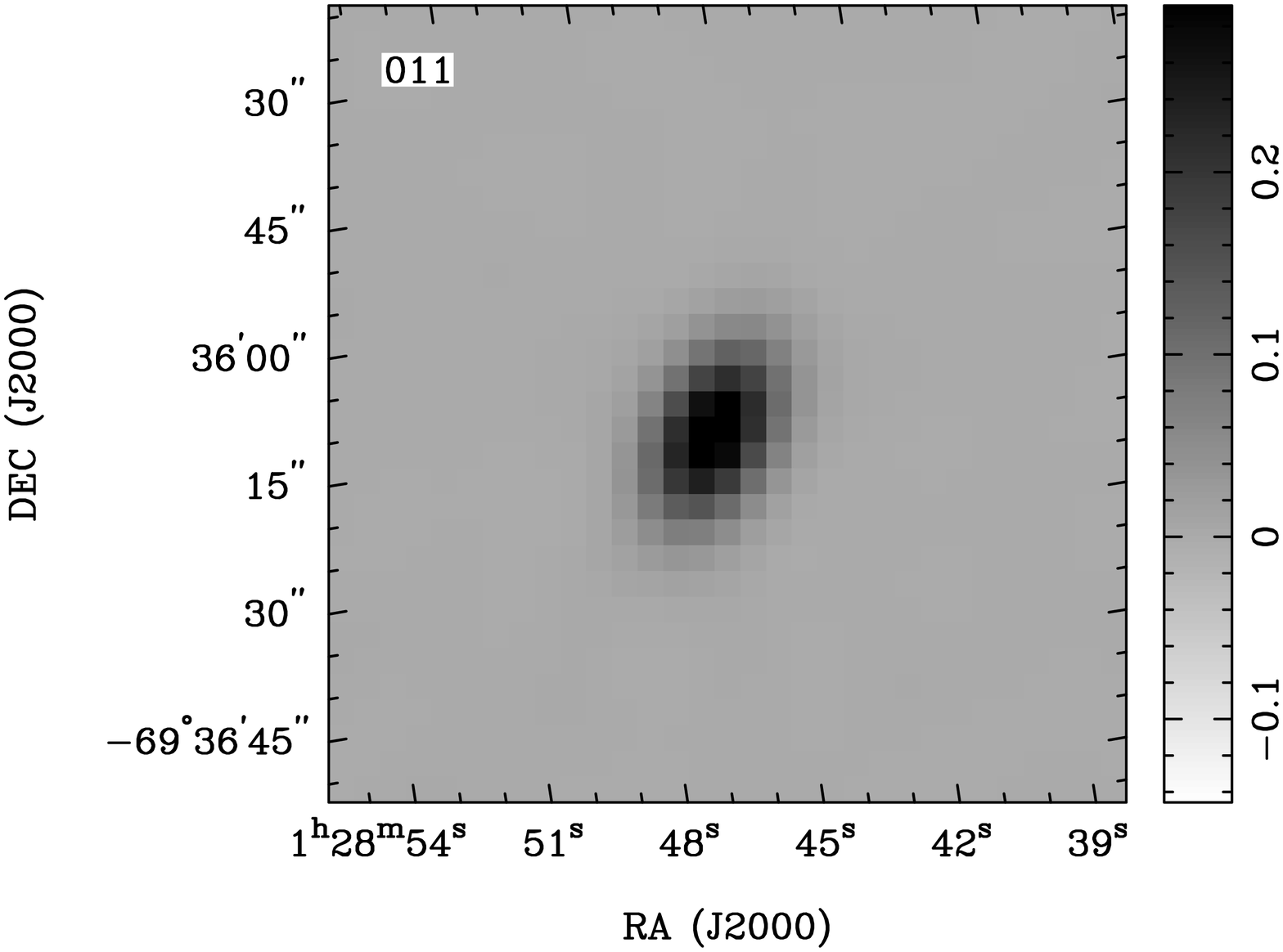}
\hspace{.05in}
\includegraphics[width=0.4\textwidth]{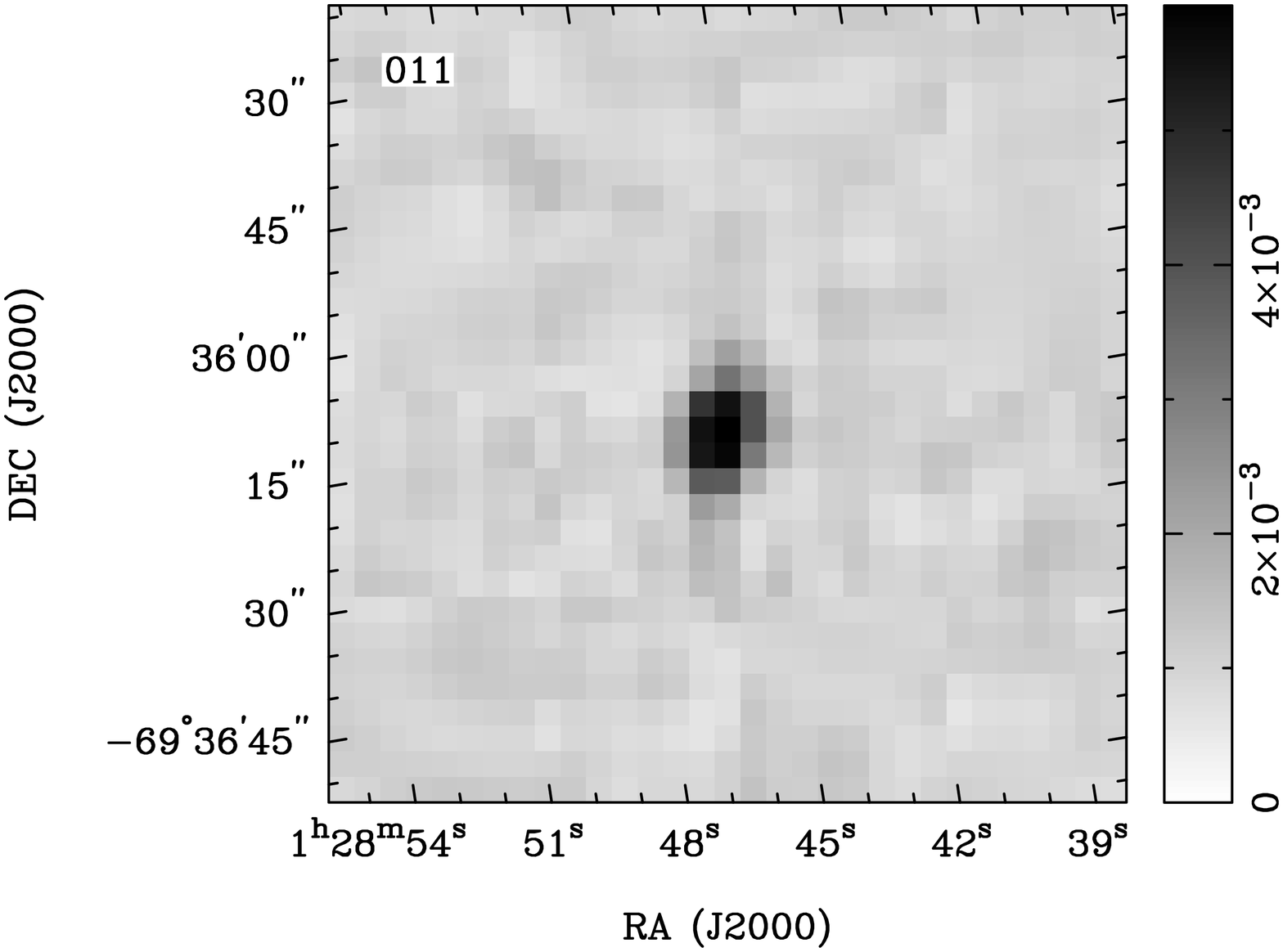}
\vspace{.05in}
\includegraphics[width=0.4\textwidth]{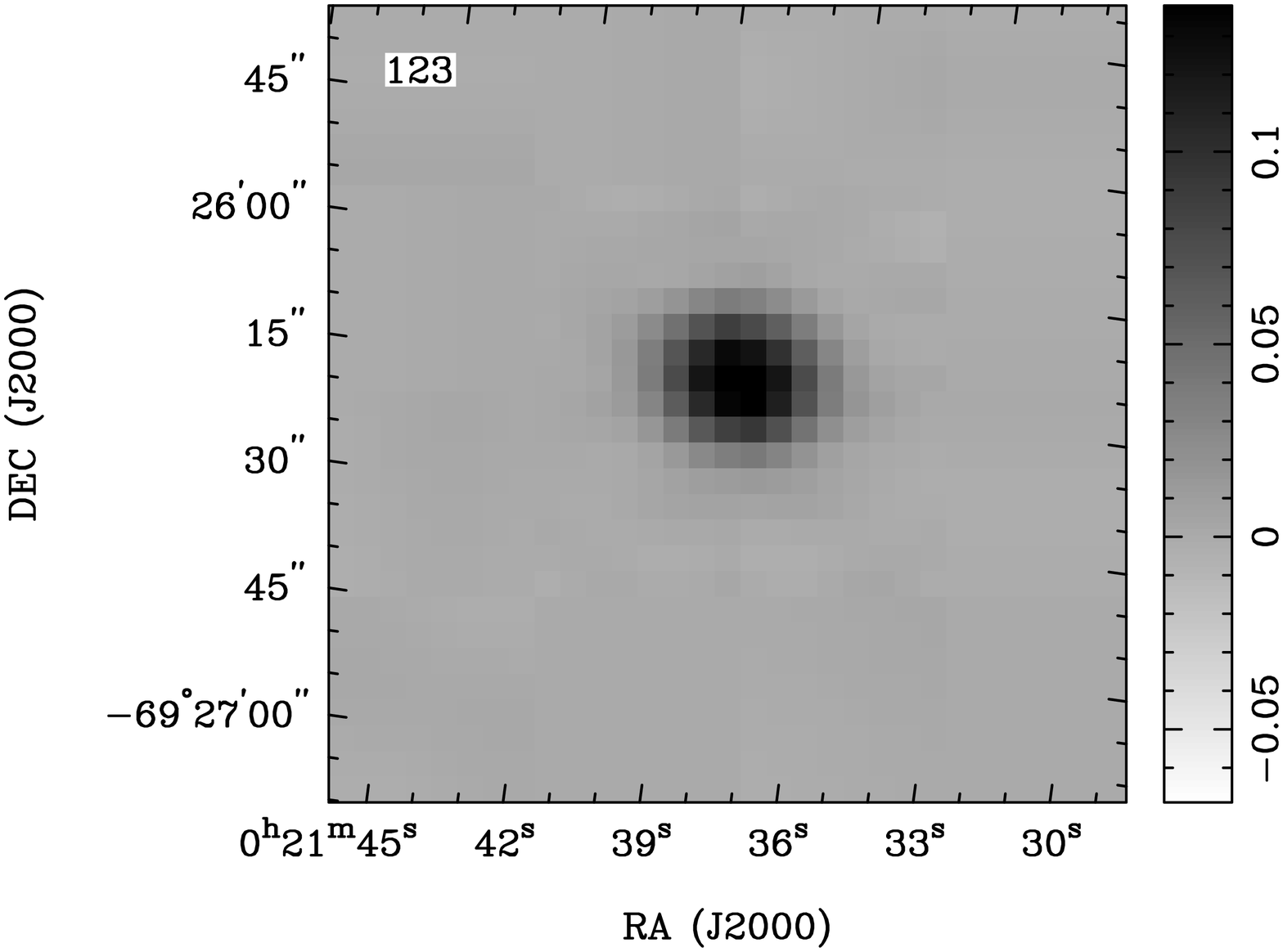}
\hspace{.05in}
\includegraphics[width=0.4\textwidth]{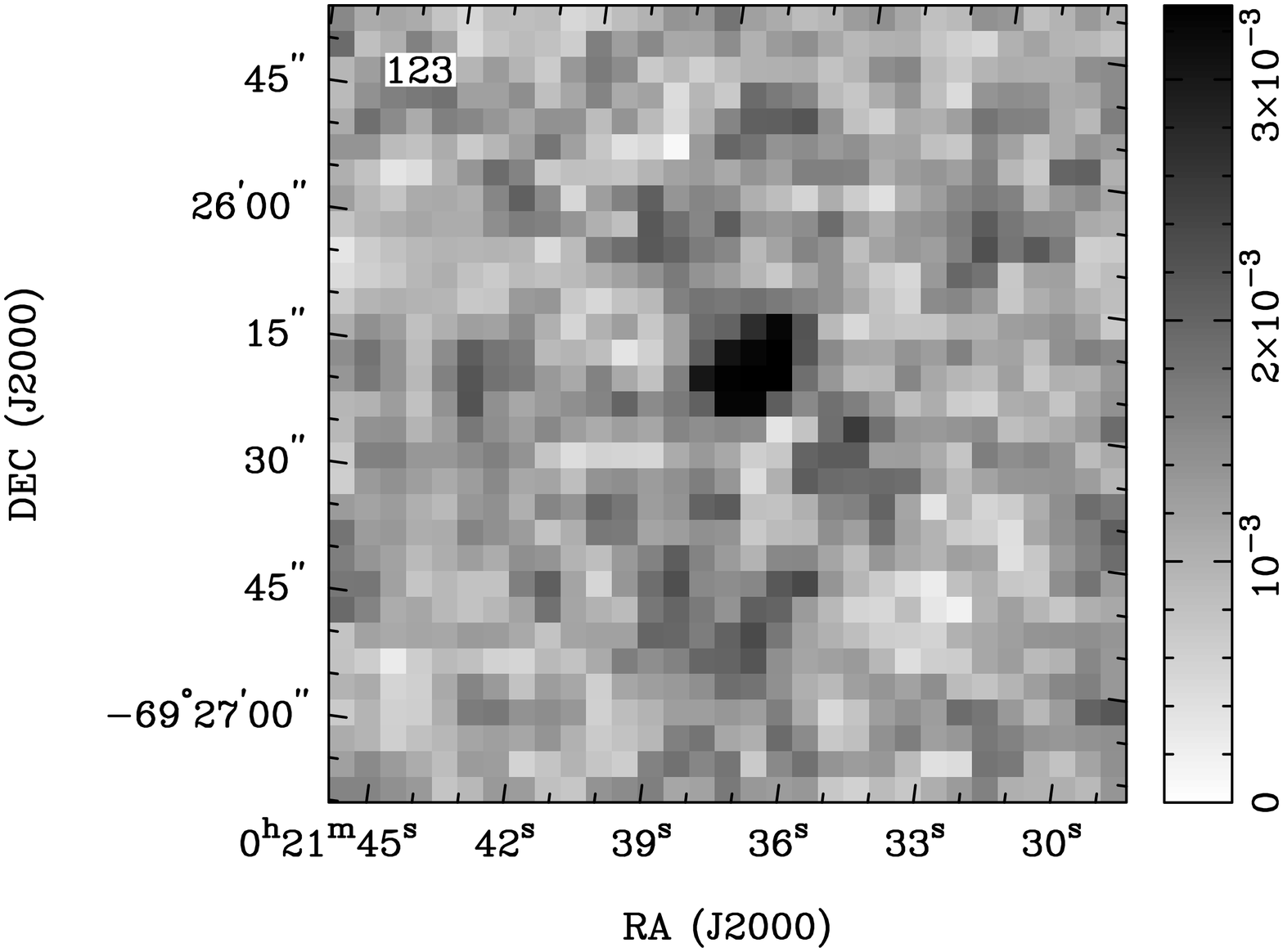}
\epsscale{1}

\caption{Two examples of polarized extragalactic background sources detected in our survey, source 011 and source 123 in Table~\ref{table:rawrms}. The left panel shows the total intensity, while the right panel shows the linear polarized intensity. Grey scale in units of Jy is shown to the left of each figure.}
\label{fig:sourceexample}

\end{figure}

For each pointing and each channel, a linearly polarized intensity (PI) map, corrected for positive bias was made. To ensure that no source was lost through bandwidth depolarization, PI maps over all channels were then averaged together to make a single polarization map for each pointing. A linearly polarized intensity map with sensitivity of 0.4 mJy per beam, covering all 440 pointings, was created using the task LINMOS. Polarized point sources were identified from the mosaicked polarized intensity image using the task SFIND, which implements the False Discovery Rate (FDR) algorithm \citep{hopkins2002}. These polarized point sources are likely to be extragalactic as their positions do not coincide with known supernova remnants (SNRs) \citep{filipovic2005}. Figure \ref{fig:sourceexample} shows two examples of linear polarization detected from extragalactic background sources in the field. For each of the 13 channel maps for each source, values of Stokes $Q$ and $U$ were extracted for the peak pixel and the eight brightest pixels (in polarized intensity) surrounding it. 

\begin{figure}
\centerline{
\includegraphics[width=0.8\textwidth]{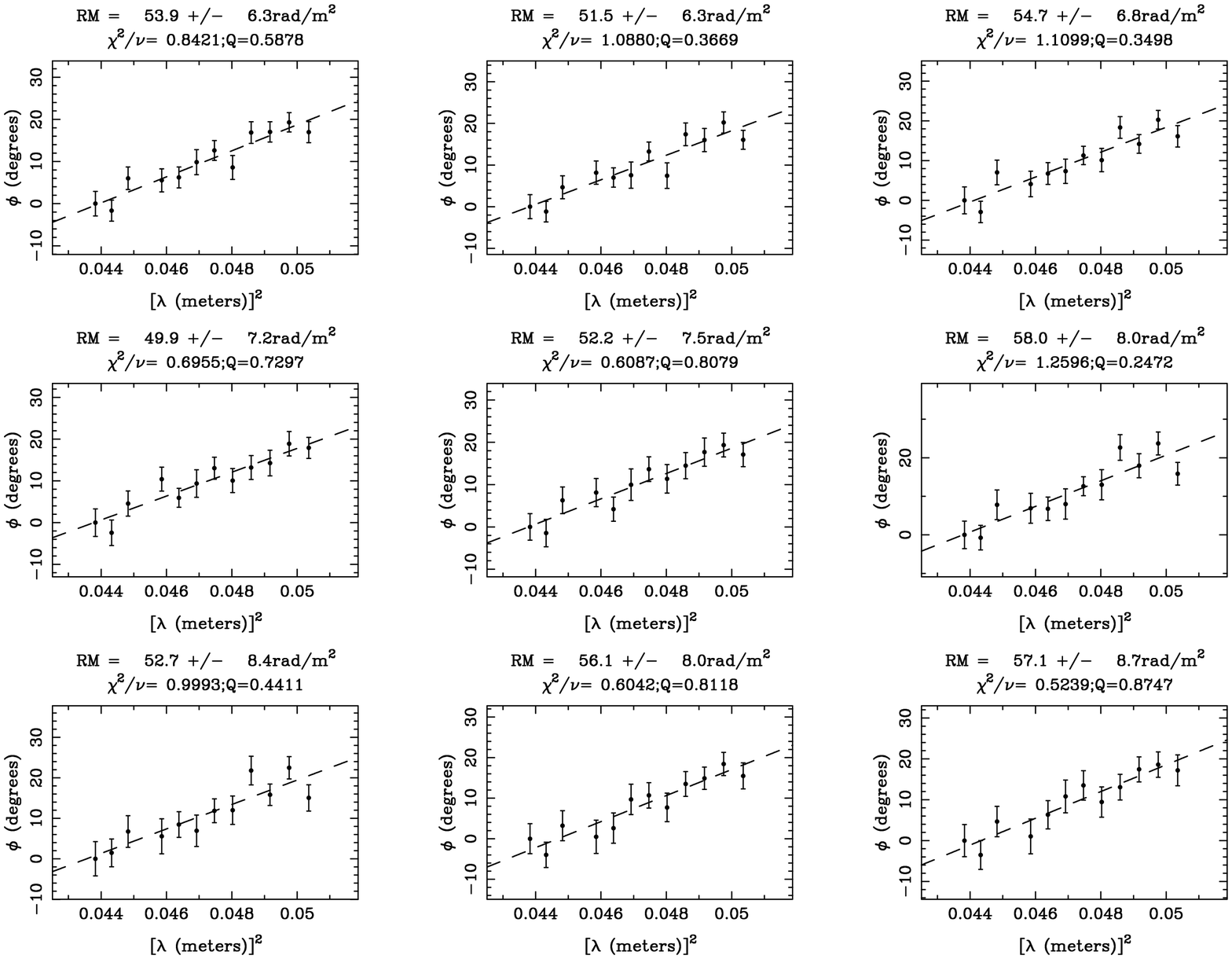}}
\caption{Least square fits of polarization angle against wavelength squared for the nine brightest pixels in polarization for extragalactic source 042 in Table~\ref{table:rawrms}. The plot at the upper left shows the fit to the brightest pixel, while the plot at the lower right corresponds to the fit of the faintest pixel. The slope of the least square fit gives the value of the RM, which is indicated above each plot. The reduced $\chi^2$ and the quality of fit (Q)  for each pixel are also displayed.}
\label{fig:rmfit}
\end{figure}

The RM of each source was computed following the algorithm developed by \cite{brown2003}. As long as the RMs have magnitudes less than $\sim$ 2700 rad m$^{-2}$, our data do not suffer from an n$\pi$ ambiguity because of the closely spaced frequency channels.  For each pixel of each source, the RM per pixel was calculated by least-squares fitting the unwrapped polarization position angle \citep[see][]{brown2003} as a function of the wavelength squared. Figure \ref{fig:rmfit} shows the least squares fit for one of our background sources. These RMs were then passed through tests to ensure sufficient signal-to-noise and a reasonable quality of fit.  A source was accepted if more than half of the pixels yield reliable RMs (quality of fit \footnotemark[2]\footnotetext[2]{The probability of a random distribution generating a value of  $\chi^2$ greater than the observed value, for $\nu$ degrees of freedom} Q $>$ 0.1) . The source RM (weighted by the error in the RM for each pixel) and its uncertainty were computed from the good pixels. If the scatter of RM from pixel to pixel within the same source was larger than twice the average statistical error of the source pixels, the source was rejected.

The data reduction procedures described above produce 70 reliable and accurate RMs as listed in Table~\ref{table:rawrms}. After comparing catalogued positions of HII regions \citep{henize1956} with those of the extragalactic background sources, we find that source 134  has a projection that coincides with N90, an active star forming region in the wing of the SMC. The RM through this particular sight line traces magnetic field and electron distribution through the HII region as well as through the diffuse ISM. 

As mentioned in \S~\ref{subsection:faradayrotation}, $\rm{RM_{Milky Way}}$ can be estimated using the RM values of extragalactic sources whose projected positions lie close to, but outside the SMC. We define the boundary of the SMC to be where the neutral hydrogen column density drops below  2$\times$10$^{21}$ atoms cm$^{-2}$ or the extinction corrected intrinsic H$\alpha$ intensity of the SMC drops below 25 deci-Rayleigh (dR), where 1~R = 10$^6$ photons per 4$\pi$ steradian = 2.42 $\times$ 10$^{-7}$ ergs cm$^{-2}$ s$^{-1}$  sr$^{-1}$ (see \S~\ref{subsection:otherdata}). A source's projection is considered to be inside the SMC if it lies inside either the HI column density or the H$\alpha$ threshold. We find that 10 extragalactic sources satisfy this criteria and are indicated with * in Table~\ref{table:rawrms}.

\begin{figure}
\centerline{
\includegraphics[width=0.8\textwidth]{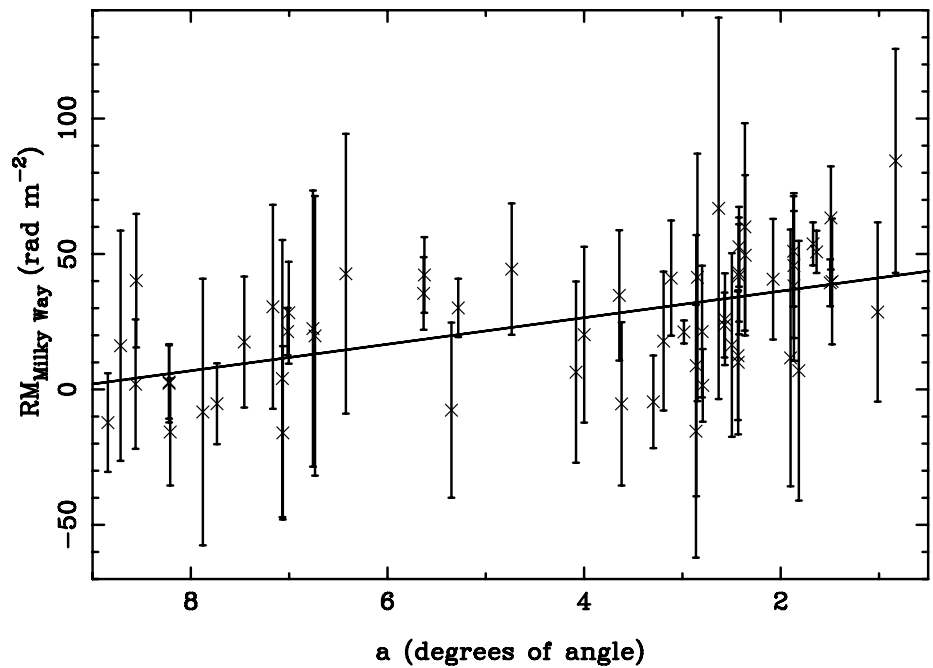}}
\caption{Foreground RM fit to the 60 extragalactic sources in Table~\ref{table:rawrms} whose projections on the sky lie outside the SMC. The foreground RM can be least square fitted as a linear function of right ascension. The best fit is ${\rm RM_{Milky Way}}$= 46.1$-$4.9$\times$a rad m$^{-2}$ with a reduced $\chi^2$ of 0.88, where a is the offset eastward from right ascension of 0 in degrees.}
\label{fig:fgrmfit}
\end{figure}

The data are insufficient to constrain a foreground RM dependence on declination as there are very few background sources at more southerly declinations. However, it is obvious that background sources to the west of the SMC have values of RM which are more positive than those to the east, hence, we perform a least square fit  to the value of the foreground rotation measure as a function of right ascension in degrees (Figure~\ref{fig:fgrmfit}). The best fit has the form
\begin{equation}
\label{eq:fit}
\rm{RM_{Milky Way}}= (46.1\pm 4.1)-(4.9 \pm 0.9)\times a~~\rm{rad~m^{-2}}
 \end{equation}
 where a is the offset in degrees eastward from zero right ascension. 
  
\begin{figure}
\centerline{
\includegraphics[width=0.8\textwidth]{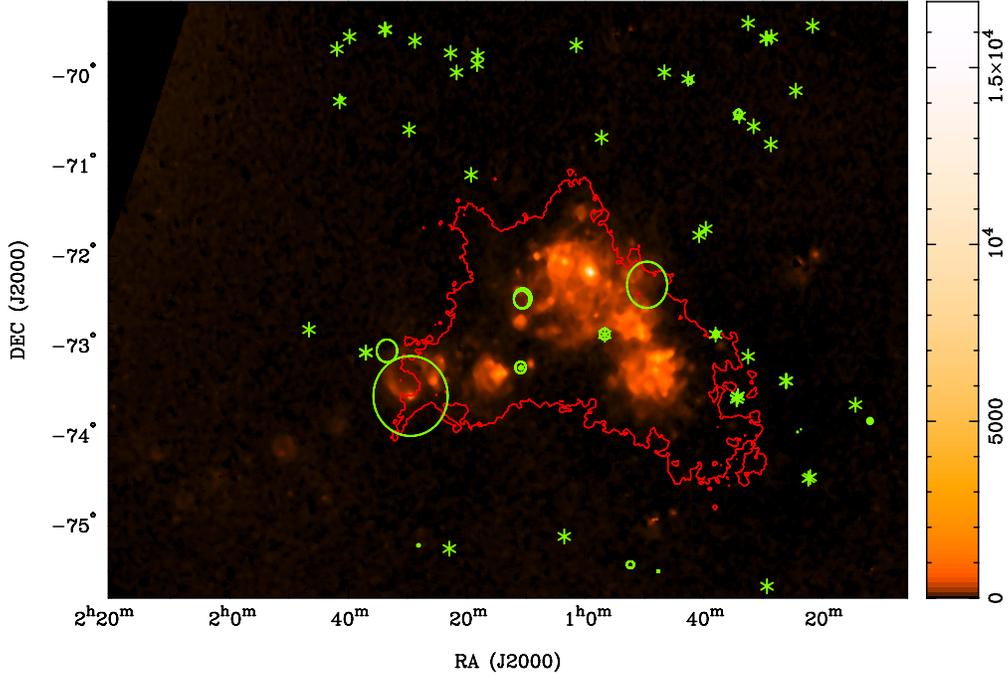}}
\caption{Distribution of foreground corrected RMs overlaid on smoothed and continuum subtracted H$\alpha$ emission from the SHASSA survey (Gaustad et al. 2001). The color scale is in units of dR. The contour represents an HI column density of 2$\times$10$^{21}$ atoms cm$^{-2}$ \citep{stanimirovic2004}. Closed and open circles represent positive and negative rotation measures respectively. Asterisks denote RMs that are consistent with zero within their uncertainties. The center of the symbol marks the position of the extragalactic source. The diameter of a circle is proportional to the value of $|$RM$|$ at that position. The largest open circle in the above figure represents a RM of $-$400 rad m$^{-2}$. The consistent pattern of negative RMs projected against the SMC indicates that the SMC has a significant coherent magnetic field along the line of sight, directed away from us.}
\label{fig:aftersubonhalpha}
\end{figure}

After subtracting the fit to the foreground RM as given in Equation~(\ref{eq:fit}) and propagating the associated uncertainties into $\rm{RM_{SMC}}$, the distribution of RM through the SMC is shown in Figure~\ref{fig:aftersubonhalpha} and listed in  Table~\ref{table:rmsources}. The RMs  of the 10 extragalactic sources which lie directly behind the SMC range from $-$400$\pm$60  rad m$^{-2}$ (source 135) to 0$\pm$50 rad m$^{-2}$ (source 136), with a weighted mean of $-$30 rad m$^{-2}$, a weighted standard deviation (calculated using Equation (4.22) in \cite{bevington2003}) of 40 rad m$^{-2}$ and a median of $-$75 rad m$^{-2}$. After the foreground subtraction, RMs of sources whose projections lie outside the SMC should be zero by construction. We find  a residual RM of 0 rad m$^{-2}$ with a weighted standard deviation of 20 rad m$^{-2}$.

From the fact that 9 out of 10 extragalactic sources behind the SMC have negative RMs and the other has a RM consistent with zero, we argue that the underlying field is unlikely to be random in direction as this would produce equal numbers of positive and negative RMs across the galaxy with a mean close to zero. If we are observing a random field, the probability of getting at least nine out of ten RMs of the same sign is 0.4\%. In other words, the magnetic field across the entire SMC is coherently directed away from us at a 99.6\% confidence level. The measured large RMs through the SMC also cast doubt on the orientation of the plane-of-the-sky magnetic field  obtained by \cite{loiseau1987} and \cite{haynes1991} from linearly polarized radio synchrotron emission, because our observed mean RM of $-$30 rad m$^{-2}$ rotates the polarization position angle by $\sim$ 70$^\circ$ at 20 cm. Since  \cite{loiseau1987} and \cite{haynes1991} did not correct for Faraday rotation, their angles do not correspond to intrinsic angles in the SMC.

\subsection{Optical Starlight Polarization Data}
\label{subsection:sldata}
\cite{mathewson1970a} observed 76 stars in the SMC, along with 60 Galactic stars towards the SMC at distances from 50 pc to 2 kpc to correct for the foreground polarization. They found that the foreground signal has a fractional polarization of 0.2\%. The distribution of SMC stars and their raw optical polarization position angles are plotted in Figure~\ref{fig:starlight}. 

As pointed out in \S~\ref{subsection:osp}, the Galactic foreground polarization is directed along the SMC-LMC connection, so a careful foreground correction is required. Schmidt (1976) subdivided the SMC's projection onto the celestial sphere into five regions and calculated the foreground correction for each region by studying a large number of Galactic foreground stars at different distances. We have applied this improved Galactic foreground correction to the 76 stars observed by \cite{mathewson1970a}. Because of the large angular extent of the SMC, measuring the deviation of polarization position angles with respect to the north is not useful. Instead, we choose our reference direction to be along the great circle joining the SMC and the LMC on the celestial sphere. The positions, polarization position angles and associated errors of the polarization vectors of 76 SMC stars, after foreground subtraction, are listed in Table~\ref{table:starlightdata}.

\subsection{H$\alpha$ and HI data}
\label{subsection:otherdata}
In order to estimate the thermal electron density in the SMC, we have used the continuum subtracted  SHASSA H$\alpha$ map of the SMC smoothed to 4 arcminutes \citep{gaustad2001}. The image has a sensitivity of 5 dR.

Knowing the H$\alpha$ intensity of the SMC and the foreground extinction allows one to evaluate the emission measure, as will be shown in \S~\ref{subsection:emissionmeasure}. To correct the observed H$\alpha$ intensity for interstellar extinction, we use the integrated neutral hydrogen (HI) column density map of the SMC presented by \cite{stanimirovic1999} from ATCA and Parkes spectral line observations. The column density was derived by integrating the 21cm HI signal over the heliocentric velocity range +90 to +215 km s$^{-1}$, and the resulting column density map has an angular resolution of 1.6 arcminutes.

\section{The Line-of-sight Magnetic Field Strength in the Warm Ionized Medium of the SMC}
\label{section:wimbfield}
In this section, we construct three ionized gas models  from the extinction corrected H$\alpha$ intensity of the SMC and from pulsar dispersion measures. These models allow us to estimate the average magnetic field strength along the line of sight, $\overline{B_{\parallel}}$, from the RMs presented in \S~\ref{subsection:radiodata}.

\subsection{Pulsar Dispersion Measure and Rotation Measure}
\label{subsection:pulsardm}
The dispersion measure (DM) of a pulsar is an integral of the electron density content along the line of sight, defined as:
\begin{equation}
\label{eq:dm}
{\rm{DM}} = \int ^{L}_{0} {n_{e}}(l){dl} = \overline{n_{e}} L ~~\rm{pc~cm^{-3}}
\end{equation}
where $\overline{{n_e}}$ is the average electron density along the total path length $L$. There 
are 5 known radio pulsars in the SMC. Their positions, measured DMs and an RM for the one source with Faraday rotation information, are listed in Table~\ref{table:radiopulsar}. 

We subtract the Galactic contribution to DMs of SMC pulsars using the NE2001 Galactic free electron model developed by \cite{cordes2002}. The average DM of pulsars in the SMC after the removal of the Galactic contribution is  $\langle \rm{DM_{SMC,pulsar}} \rangle $ $=$ 80.9 pc cm$^{-3}$. If we assume that the pulsars are evenly distributed through the SMC, the total DM through the SMC is approximately twice the mean value, that is, $\rm{\langle DM_{SMC} \rangle}$~$\approx$~162 pc cm$^{-3}$. 

Following the treatment of \cite{manchester2006}, the mean electron density $\langle$$n_e$$\rangle$ in the SMC can be estimated by computing the dispersion of DMs and the dispersion of pulsars' spatial coordinates. The underlying assumption is that the SMC is spherically symmetric. We assume that the mean distance to the SMC pulsars is 60 kpc, instead of 50 kpc as in \cite{manchester2006}, and that the offsets of pulsar locations in RA and DEC directions are independent. The mean electron density in the SMC is given by
\begin{equation}
\langle n_{e} \rangle = \frac {\rm{\sigma_{DM}}} {\rm{\sigma_{spatial,1D}}}
\end{equation}
where $\rm{\sigma_{DM}}$~$\approx$~48 pc cm$^{-3}$ is the dispersion of pulsar DMs, after foreground subtraction;  and $\rm{\sigma_{spatial,1D}}$~$\approx$~1230 pc is the one-dimensional spatial dispersion of their positions. This estimation gives a mean electron density of $\langle$n$_e$$\rangle$~$\approx$ 0.039~cm$^{-3}$ in the SMC. 

Out of the 5 known radio pulsars in the SMC, only one (PSR J0045-7319) has a measured rotation measure, with a value of $-$14 $\pm$ 27 rad m$^{-2}$ \citep{crawford2001}.  Comparing the DM of this pulsar from Table~\ref{table:radiopulsar} with $\rm{\langle DM_{SMC} \rangle}$, one can conclude that this pulsar is located approximately half way through the galaxy. Following the foreground subtraction procedure described in \S~\ref{subsection:radiodata}, the component of the RM from this pulsar that results from the magnetized medium in the SMC is $-$40~$\pm$ 30 rad m$^{-2}$. This negative value is consistent with negative signs of RMs of extragalactic sources through the SMC as given in Table~\ref{table:rmsources}.

\subsection{Emission Measure}
\label{subsection:emissionmeasure} 
\noindent The emission measure (EM) of ionized gas along the line of sight is defined as
\begin{equation}
\label{eq:em}
{\rm{EM}} = \int ^{L}_{0} {n_{e}(l)^2 }{dl} = \overline{n_{e}^2} L~~\rm{pc~cm^{-6}}
\end{equation}
where $\overline{n_e^2}$ is the average of the square of the electron density along the total path length $L$.

We derive an emission measure map of the SMC from the smoothed and star-subtracted H$\alpha$ emission in this region \citep{gaustad2001} by correcting for both foreground extinction caused by dust in the Milky Way and internal extinction in the SMC. The foreground Milky Way contribution to the observed H$\alpha$ emission is estimated by the off source H$\alpha$ intensity in regions surrounding the SMC. We assume a constant Galactic foreground HI column density of (4.3 $\pm$ 1.3)$\times$10$^{20}$ atoms cm$^{-2}$ \citep{schwering1991} and a dimensionless Galactic dust-to-gas ratio  $k$ of 0.78 \citep{pei1992}, where $k$ is defined as:
\begin{equation}
k \equiv 10^21(\tau_{\rm B}/N_{\rm HI})~~~~~~{\rm cm^{-2}}
\end{equation}
where $\tau_B$ denotes the optical depth in the optical $B$ band and $N_{\rm HI}$ denotes the neutral hydrogen column density. For the internal extinction of the SMC, the correction is derived from the HI column density map \citep{stanimirovic2004} and a dust-to-gas ratio $k$ of 0.08 \citep{pei1992}.  We have used the empirical extinction curves of the Milky Way and the SMC at the wavelength of H$\alpha$ ( $\lambda_{\rm H\alpha}$ = 6563\AA)
 \begin{equation}
\xi(\lambda_{\rm H\alpha})=\tau_{\rm H\alpha}/ \tau_{\rm B} = 0.6
\end{equation}
\citep{pei1992}, where $\tau_{\rm H\alpha}$ is the optical depth at 6563\AA. The optical depth at the wavelength of H$\alpha$ can thus be expressed as
  \begin{equation}
\tau_{\rm H\alpha} = k (N_{\rm HI}/ (10^{21} {\rm{cm^{-2}}} ))  \xi(\lambda_{\rm H\alpha})
 \end{equation} 
The intrinsic H$\alpha$ intensity of the SMC is calculated assuming\footnotemark[3] \footnotetext[3]{See Appendix~\ref{appendix:hacorrection} for details} that the H$\alpha$ emitting gas is uniformly mixed with dust in a region of optical depth $\tau_{\rm H\alpha}$. The EM for H$\alpha$ intensity $I_{\rm H_{\alpha},intrinsic,SMC}$ produced by gas at electron temperature $T_e$ is \cite[see for example][]{lequeux2005}:

\begin{equation}
\label{eq:Haemconversion}
{\rm{EM}}=  \frac{{I_{\rm{H_\alpha,intrinsic,SMC}}} (T_e/10000{\rm K})^{0.5}} {0.39(0.92-0.34\ln(T_e/10000K))}
\end{equation}
where T$_{e}$ is the electron temperature of the diffused ionized medium in the SMC and $I_{\rm H_{\alpha},intrinsic,SMC}$ in Rayleighs is the intrinsic H$\alpha$ intensity of the SMC .

\begin{figure}
\centerline{
\includegraphics[width=0.8\textwidth]{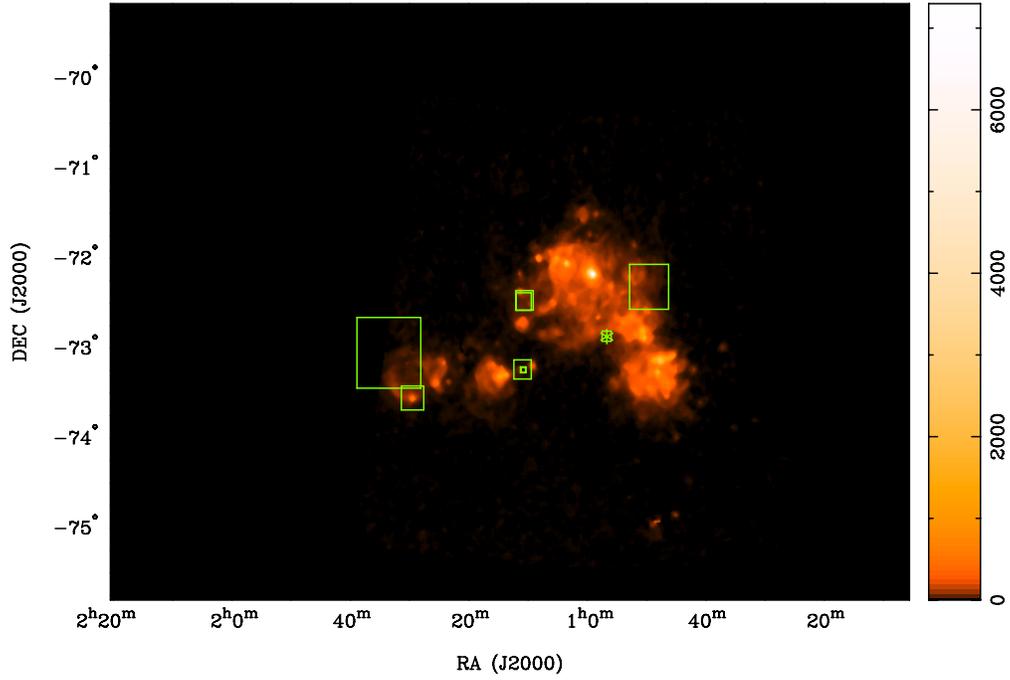}}
\caption{The line-of-sight magnetic field strength through the SMC derived using ionized gas model 3 ( \S~\ref{subsubsection:model3}). The image is the extinction corrected emission measure map of the SMC, in units of pc cm$^{-3}$ (as shown by the scale bar to the right of the image), derived from the SHASSA H$\alpha$ survey. Open squares denote magnetic fields whose line of sight component direct away from us while the asterisk denotes a magnetic field strength consistent with
 zero within one standard deviation. The centre of the squares mark the positions of the extragalactic sources. The length of a side of the square is proportional to the line of sight magnetic field strength. The largest open square in the above figure represents a field strength of  $-$2.0 $\mu$G. This figure illustrates that the SMC hosts a large scale coherent magnetic field of the order of $\sim$ 0.2 $\mu$G.}
\label{fig:boverlayem}
\end{figure}

 As no measurement of the temperature of SMC's diffuse ionized medium exists in the literature, we estimate $T_e$ by adding 2,000K to the average temperature in HII regions ($\sim$ 12,000K \citep{dufour1977}) in the SMC, by analogy with the diffused ionized medium in the Milky Way, which are $\sim$ 2,000K hotter than Galactic HII regions \citep{madsen2006}. We thus adopt T$_{e}$ $\sim$ 14,000 K. The resulting emission measure map is shown in Figure~\ref{fig:boverlayem}.

\subsection{Diffuse Ionized Gas Models} 
\label{subsection:ionizedgasmodels}

Our models are based on the assumption that there is no correlation between the fluctuations in the electron density and in the magnetic field. From Equation~(\ref{eq:rmdef}) the average magnetic field strength along the line of sight $\overline{B_{\parallel}}$ is then:
\begin{equation}
\label{eq:solveb}
\overline{B_\parallel} = \frac { \rm{RM_{SMC}} } {0.812~\overline{ n_e}L}
\end{equation}

For gas densities lower than 10$^3$ cm$^{-3}$, there is no observational evidence of correlation between the magnetic field strength and gas density \citep{crutcher2003}. As discussed by \cite{beck2003}, if pressure equilibrium is maintained, one expects an anti-correlation between the magnetic field and the thermal electron density on small scales. This would lead to an underestimation of $\overline{B_{\parallel}}$. On the other hand, $n_e$ and ${B_{\parallel}}$ might be correlated by compression in SNR shocks in selected regions. This would lead to an overestimation of $\overline{B_{\parallel}}$. Depending on the property of the turbulent ISM in the SMC, our estimates of $\overline{B_\parallel}$ presented in the following are potentially subject to systematic bias by a factor of two to three.

The following models ignore the presence of individual HII regions and only focus on diffused ionized regions. As mentioned in \S~\ref{subsection:radiodata}, source 134 appears to coincide with N90, an active star formation region in the SMC. Estimations of $\overline{B_{\parallel}}$ in the following models along this particular sight line might not reflect the true value and it is excluded when calculating the average line-of-sight magnetic field strength of the SMC, $\langle B_{c,\parallel} \rangle$.

\subsubsection{Model 1: Constant Dispersion Measure}
\label{subsubsection:model1}
We can estimate the line of sight magnetic field strength ($\overline{B_{\parallel}}$) by assuming that the dispersion measure through the SMC is constant across the galaxy, with $\rm{\langle DM_{SMC} \rangle}$=$\overline{n_e}$ $L$ = 162 pc cm$^{-3}$. Combining Equations~(\ref{eq:dm}) and (\ref{eq:solveb}), we find
\begin{equation}
\label{eq:model1}
\overline{B_{\parallel}}  = \frac {\rm {RM_{SMC}}} {0.812 \rm{\langle DM_{SMC} \rangle}}
\end{equation}

Estimations of $\overline{B_{\parallel}}$  obtained using this model are listed in the second column of Table~\ref{table:b}. The weighted magnetic field along the line of sight averaged across the SMC is $-$0.20 $\mu$G. This model gives crude estimates of  $\overline{B_{\parallel}}$, since in reality, both the depth of the SMC ($L$) and the line of sight average of electron density $\overline{ n_e}$ vary from one sight line to the other, while their product need not stay the same.

\subsubsection{Model 2: Constant $\overline{ n_e}$, varying L}
\label{subsubsection:model2}

The motivation behind this model is the observational evidence for the variation of the line-of-sight depth as a function of position in the SMC, as seen in both the HI velocity dispersion \citep{stanimirovic2004} and the variation of the distance modulus to Cepheid variables \citep{lah2005} across the SMC.

It is useful to define filling factor $f$, the fraction of the total path length occupied by thermal electrons, as the following \cite[see for example][]{reynolds1991,berkhuijsen2006}
\begin{equation}
\label{eq:ffdef1}
f=\frac{\overline{{n_e}}}{\overline{n_{\rm cloud}}}
\end{equation}
where $\overline{n_{\rm cloud}}$ is the average electron density in ionized gas clouds along the line of sight and $\overline{ n_e}$  is the mean electron density along the line of sight. For the special case where the electron densities in the individual ionized gas clouds along the line of sight are the same, the filling factor can be expressed as
\begin{equation}
\label{eq:ffdef2}
f = \frac{\overline{ n_e}^2} {\overline{{n_e}^2}}
\end{equation}

In this model, we assume that the filling factor $f$, and the mean electron density along the line of sight $\overline{ n_e}$ remain the same across the galaxy, while the depth $L$ of the SMC varies. Additionally, we assume that all ionized gas clouds along the line of sight have the same electron density $n_{\rm cloud}$. Using the definition of the filling factor $f$ given in Equations~\ref{eq:ffdef1} and \ref{eq:ffdef2}, one can express the average DM and EM through the SMC as
\begin{equation}
\label{eq:dmnc}
{\rm \langle DM_{SMC} \rangle} = \overline{ n_e} {\langle L_{\rm SMC} \rangle}= \overline{n_{\rm cloud}} f {\langle L_{\rm SMC}\rangle} = n_{\rm cloud} f {\langle L_{\rm SMC} \rangle}
\end{equation}
\begin{equation}
\label{eq:emnc}
{\rm \langle EM_{SMC} \rangle} = \overline{ n_e^2} {\langle L_{\rm SMC}\rangle} = \overline{n_{\rm cloud}^2} f  {\langle L_{\rm SMC}\rangle} =n_{\rm cloud}^2  f  {\langle L_{\rm SMC} \rangle}
\end{equation}
where ${\langle L_{\rm SMC} \rangle}$ denotes the average depth of the SMC. Combining the two equations above, we obtain
\begin{equation}
{n_{\rm cloud}} = \frac  {\rm{\langle {EM_{SMC}}\rangle}} {\rm{{\langle DM_{SMC} \rangle}}}
\end{equation}

We assumed in the previous sections that $\rm{\langle DM_{SMC} \rangle}$ = 162 pc cm$^{-3}$. $\rm{\langle EM_{SMC} \rangle}$ is estimated by the average emission measure through the SMC defined by the HI column density contour in Figure~\ref{fig:aftersubonhalpha}. Individual HII regions with intrinsic H$\alpha$ intensity higher than 500 dR are masked out before taking the average, resulting in $\rm{\langle EM_{SMC} \rangle} \approx$ 16 pc cm$^{-6}$. This yields a mean density in an ionized cloud $n_{\rm cloud}$ $\approx$ 0.10 cm$^{-3}$. The filling factor $f$ is estimated using Equation ~\ref{eq:ffdef1} assuming a mean electron density $\overline{ n_e}$ = $\langle$$  n_e$$\rangle$ $\approx$  0.039 cm$^{-3}$ obtained from the pulsar DM analysis in \S~\ref{subsection:pulsardm}, to yield $f=0.39$. Using the definition of filling factor given in Equation~\ref{eq:ffdef2}, we can express the EM towards an extragalactic background source as
\begin{equation}
{\rm EM_{source}} = n_{\rm cloud} ^2 f {L_{\rm source}}
\end{equation}
Solving for $L_{source}$, the path length through the SMC to an extragalactic source, yields
\begin{equation}
L_{\rm source}= \frac {\rm{EM_{source}}} { f {n_{\rm cloud}}^2}
\end{equation}
Using the above equation, we have computed the average path length through the SMC to the extragalactic sources to be $\langle L \rangle$ $\approx$ 10 kpc with a standard deviation $\Delta L$ $\approx$ 6 kpc, which is roughly consistent with \cite{stanimirovic2004}.

Using additional information from H$\alpha$, the mean magnetic field strength parallel to the line of sight can be found by
\begin{equation}
\overline{B_{\parallel}} =  \frac {\rm {RM_{SMC}}}{0.812 \rm{\langle DM_{SMC} \rangle }}\left(\frac {\rm{\langle EM_{SMC} \rangle }} {\rm{EM_{source}}} \right)
\label{eq:model2}
\end{equation}
This equation has a similar form to Equation~(\ref{eq:model1}) with an additional correction factor for the variation of EM across the galaxy. Values for $\overline{B_{\parallel}}$ using this estimate are listed in the third column of Table~\ref{table:b}. The weighted magnetic field along the line of sight averaged across the SMC is $-$0.16 $\mu$G.
 
\subsubsection{Model 3: Constant Occupation Length $fL$}
\label{subsubsection:model3}
In this model, we assume that the occupation length $fL$ (the effective length occupied by thermal electrons along a sight line) is constant while the line-of-sight mean electron density $\overline{n_e}$ is allowed to vary between sight lines. In addition, we assume that $n_{\rm cloud}$, the density of all ionized clouds along the line of sight, is the same. One can manipulate Equations~(\ref{eq:dmnc}) $\&$ (\ref{eq:emnc}) and solve for the occupation length $fL$:

\begin{equation}
fL = \frac {\rm{\langle DM_{SMC} \rangle}^2} {\rm{\langle EM_{SMC} \rangle}}
\end{equation}
We found in \S~\ref{subsection:pulsardm} that $\rm{\langle DM_{SMC} \rangle}$ $=$ 162 pc cm$^{-3}$.  $\rm{\langle EM_{SMC} \rangle }$ $\approx$ 16 pc cm$^{-6}$  was estimated in \S~\ref{subsubsection:model2}. This yields an occupation length $fL$ $\approx$ 1.6 kpc.

Using the definition of filling factor given in Equation~\ref{eq:ffdef2}, we can express EM towards an extragalactic background source as
\begin{equation}
{\rm EM_{source}}=\overline{ n_e^2} L = \overline{n_e}^2 L/  f
\end{equation}
Solving for $\overline{n_e}$ yields
\begin{equation}
\overline{ n_e} = \sqrt { \frac{ {\rm EM_{source}}f } {L}}
\end{equation}
Substituting the above expressions into Equation~(\ref{eq:solveb}) and solving for the mean magnetic field strength through different sight lines gives
\begin{equation}
\overline{B_{\parallel}} = \frac { \rm{RM_{SMC}}}{0.812  \rm{\langle DM_{SMC} \rangle}} \sqrt{ \frac {\rm{\langle EM_{SMC} \rangle}} {\rm{EM_{source}}}}
\label{eq:model3}
\end{equation}
This equation has a similar form to Equation~(\ref{eq:model2}), but with a square root rather than linear dependence on EMs. Values of $\overline{B_{\parallel}}$ derived using this method are listed in the fourth column of Table~\ref{table:b}. The weighted magnetic field along the line of sight averaged across the SMC is  $-$0.19 $\mu$G.

\subsubsection{Summary of the Models}
\label{subsubsection:modelsummary}
All models yield field strengths that are mostly consistent with each other within their uncertainties. Model 1 is a simplified picture of the physical situation which does not make use of all the known information, and thus it provides rough estimates of $\overline B_{\parallel}$. Model 2 and 3 are more sophisticated, and thus they provide estimates that probably better describe the  true $\overline B_{\parallel}$. Out of the three models, model 2 makes use of the most information one can get from pulsar dispersion measure analysis and the H$\alpha$ intensity of the SMC. Model 3 has the most degrees of freedom: both $f$ and $L$ are allowed to vary from one line of sight to the other as long as their product stays the same, and the average electron density along the line of sight $\overline{n_e}$ is allowed to change depending on which sight line one is looking through. However, one should note that both model 2 and 3 assume that the electron density in ionized clouds along the line of sight is either $n_{\rm cloud}$ (a constant) or 0, that is, a smooth fluctuation of electron density along the line of sight is forbidden. The following discussions refer to estimations from model 3 unless specified otherwise. 
 
The 10 extragalactic sources that lie behind the SMC yield line-of-sight magnetic field strengths ranging from $-$1.5 $\pm$ 0.6 (source 135) to 0.0 $\pm$ 0.2 $\mu$G (source 136), where the negative sign denotes a magnetic field directed away from us. Only three out of the 10 sources have a $\overline B_{\parallel}$ consistent with zero, while all the others are negative at at least the 2$\sigma$ and four are negative at the 3 $\sigma$ confidence level. The distribution of the magnetic field through the SMC is plotted in Figure~\ref{fig:boverlayem} on the emission measure map of the galaxy. The weighted mean of the line-of-sight strength of the magnetic field is $-$0.19 $\mu$G, which in subsequent discussion we adopt as the coherent magnetic field strength of the SMC parallel to the line of sight, $\langle B_{c,\parallel} \rangle$.



Using the standard error in the weighted mean prescription in \cite{cochran1977} and assuming that there is one overall underlying field in the SMC; we find that $\langle B_{c,\parallel} \rangle$ = $-$0 .19 $\mu$G with a standard error in the weighted mean of 0.06 $\mu$G. We can also quantify the scatter of the field by quoting at 68 $\%$ confidence level that the coherent magnetic field strength of the SMC, $\langle B_{c,\parallel} \rangle$ is $-0.2 ^{+0.1} _{-0.3}$ $\mu$G. Note that the strength of the magnetic field that we derive in this section is independent of the temperature of the WIM one picks to convert the H$\alpha$ intensity into an EM (see Equation~\ref{eq:Haemconversion}) because the expressions of the field strength (Equation~\ref{eq:model1}, \ref{eq:model2} $\&$ \ref{eq:model3}) only involve the ratio of emission measures.

\section{The Plane-of-the-sky Magnetic Field of the SMC}
\label{section:posb}
\subsection{Estimation of $\langle B_{o,\perp} \rangle $using the C-F method}
\label{subsection:starlightcf}

In \S~\ref{section:wimbfield}, we have derived the line-of-sight magnetic field strength of ionized gas in the SMC using the RMs of extragalactic radio sources. Now, we would like to estimate the strength of the magnetic field perpendicular to the line of sight by applying the C-F method to the starlight polarization data presented in \S~\ref{subsection:sldata}. Since the infrared dust emission of the SMC has a similar morphology to its H$\alpha$ intensity, we assume that dust is in the warm ionized medium  (WIM) of the SMC. In addition, \cite{rodrigues1997} found, from analysis of extinction and polarization data of SMC stars, that the SMC has smaller grain sizes than those in the Milky Way (where dust lie mostly in the warm neutral medium (WNM), see \cite{lockman2005}). One expects smaller dust grains in the WIM than in the WNM due to grain shattering and grain-grain collisions \citep{jones1996}. The fact that the SMC has smaller grain sizes than those in the Milky Way further supports our assumption that dust is in the WIM of the SMC. To estimate the strength of the plane-of-the-sky component of field, we use Equation~(\ref{eq:cfmethod}). The density of the medium, $\rho$, is
\begin{equation}
\rho=\gamma_Hn_Hm_H
\end{equation}
where $\gamma_H$ $\approx$ 1.22 is the equivalent molecular weight of the ISM for SMC abundances \citep{russell1992,peimbert2000}, $n_H$ is the number density of hydrogen and $m_H$ is the mass of a hydrogen atom. We assume that at a temperature of 14,000K, hydrogen in the WIM is completely ionized, with a negligible ionization fraction for heavier elements. Hence, the mean ISM hydrogen number density is $n_H$ = $n_{\rm cloud}$  $\sim$ 0.1 cm$^{-3}$ (see \S~\ref{subsubsection:model2}). The average HI line-of-sight velocity dispersion is 22 $\pm$ 2 km s$^{-1}$ \citep{stanimirovic2004}, which we adopt as the line-of-sight velocity dispersion in the WIM also. 

To estimate the ordered field strength in the plane of the sky, we eliminate stars with large uncertainty in the polarized fraction as well as stars that lie outside the main body of the SMC as defined by the column density contour value of 2$\times$10$^{21}$ atoms cm$^{-2}$ (see Figure~\ref{fig:aftersubonhalpha}). The starlight polarization measurements that are used in this calculation are listed in Table~\ref{table:starlightdata}.  The average polarization position angle $\langle \theta_{p} \rangle$ deviates  $+$4$^{\circ}$$\pm$12$^{\circ}$ from the great circle joining the SMC to the LMC, measured counterclockwise. The standard deviation of tan$\delta_p$ is then calculated. Using Equation~(\ref{eq:cfmethod}), the ordered component of the magnetic field in the plane of the sky averaged over the whole SMC is $\langle B_{o,\perp} \rangle$ = 1.6 $\pm$ 0.4 $\mu$G. We choose not to break the SMC up into sub-regions and estimate the plane-of-the-sky field in each because there are not a sufficient number of polarization measurements in each sub-region to sensibly estimate $\langle \theta_{p} \rangle$ and  $\sigma(tan \theta_p)$. This analysis is complementary to the rotation measure study since it provides information on the ordered component of the plane-of-the-sky magnetic field. We assume that the fields obtained using the C-F method and the RM method are orthogonal components of the same large scale field, so for a magnetic field whose line-of-sight component is coherent and whose plane-of-the-sky component is ordered, the 3D magnetic field vector is likely to be coherent as well. Hence, we write $\langle B_{c,\perp} \rangle $  = $ \langle B_{o,\perp} \rangle$.

However, this calculation is subjected to various uncertainties. In the Milky Way, it is found that dust exists mainly in the WNM \citep{lockman2005} but WIM dust emission has also been detected \citep{lagache1999}. In the above calculation, we have assume that all dust lie in SMC's WIM. In reality, some dust must be present in the WNM of the SMC (due to the correlation of IR dust emission and HI column density, see \cite{stanimirovic2000}) and hence when estimating $\rho$ and $\sigma_{v_{los}}$ in the SMC, one should take into account of the contribution from the neutral medium as well as the ionized medium. Also, if the polarization measurements are from stars on the near side of the SMC, we are merely probing the ``surface'' magnetic field of the galaxy \citep{magalhaes1990}. Furthermore, since dust regions entrenched in oppositely directed magnetic fields would polarize starlight in the same fashion, the plane-of-the-sky magnetic field strength derived is correct only if the ordered magnetic field direction does not change appreciably along the entire line of sight. We will overestimate the field strength when the plane-of-the-sky field reverses direction along the line of sight. A correction factor was introduced by \cite{myers1991} to account for this effect. We do not need to correct for this here if the C-F method and the RM method probe the same large scale field, since RM data demonstrate that the field does not reverse on large scales along the line-of-sight.

 \subsection{Estimation of $\langle B_{total,\perp} \rangle$}
 \label{subsection:equipartition}
We now compute the total plane-of-the-sky (i.e. random and ordered fields combined) magnetic field strength  $\langle B_{total,\perp} \rangle$ using equipartition energy arguments. If we assume that the cosmic ray energy density is the same as the magnetic field energy density, one can estimate $\langle B_{total,\perp} \rangle$ using the relations given in \cite{pacholczyk1970} and \cite{melrose1980} between the specific intensity of synchrotron emission, the total plane-of-the-sky field and the synchrotron emitting path length through the galaxy. We assume that the synchrotron emitting layer of the SMC has the same thickness as the Faraday rotating layer, i.e.  $L_{\rm synchrotron}$ = $\langle L \rangle$ $\approx$ 10 kpc (\S~\ref{subsubsection:model2}). Using a spectral index $\alpha$ = 0.87, a cosmic ray energy density $K$= 5$\times$ 10$^{-17}$ erg$^{-1}$ cm$^{-3}$ \citep{beck1982}, and a non-thermal intensity $I_\nu$ = 6.4 $\times$ 10$^{-20}$ erg s$^{-1}$ cm$^{-2}$ Hz$^{-1}$ sr$^{-1}$ at $\nu$ = 2.3 GHz \citep{loiseau1987}, we obtain $\langle$ $B_{total,\perp}$ $\rangle$ $\approx$ 2.2 $\mu$G.

As pointed out by \cite{beck2005}, the above calculation is likely to underestimate $\langle B_{total,\perp} \rangle$ due to the uncertainty in $K$, and the integration of the radio spectrum over a fixed frequency range (instead of a fixed energy range to approximate the cosmic ray spectrum). We have used the revised equipartition estimate of the magnetic field given in \cite{beck2005} to compute $\langle B_{total,\perp} \rangle$. Using the ratio of number densities of protons to electrons for cosmic rays accelerated in SNRs $K_0$ $\approx$ 100 and $L_{\rm synchrotron}$ $\approx$ 10 kpc, we obtain an equipartition field strength $\langle B_{total,\perp} \rangle$ = 3.2 $\mu$G.

\subsection{The Random Magnetic Field in the SMC}
\label{subsection:randomB}

The random components of the magnetic fields in the Milky Way and in the LMC are found to dominate over the ordered components \citep{beck2000,gaensler2005}. From the dispersion of RMs in the SMC, one can estimate the strength of the random component of the magnetic field\footnotemark[5]\footnotetext[5]{If there is no random field and the uniform component is coherent throughout the galaxy, there will still be an RM gradient across the galaxy due to projection onto the curved celestial sphere. We ignore this small effect and assume that the patch of celestial sphere towards the SMC is flat.}. 

To allow comparison of the random field derived by combining the synchrotron intensity and starlight polarization measurements (see the next paragraph), which has the same assumptions as ionized gas model 2 in \S~\ref{subsubsection:model2}, we construct the random magnetic field model of the SMC based on the same ionized gas model. We assume that the average electron density along the line of sight, $\overline{n_e}$, is the same through all lines of sight but that the depth of the SMC, $L$, changes from one sight line to another. We decompose the magnetic field along each sight line into coherent and random components, such that the coherent component does not vary across the SMC; the differences between the magnetic field strengths along different sight lines are only due to the random component. In Appendix~\ref{appendix:randomB} we show that the corresponding dispersion in RM is:
\begin{equation}
\label{eq:rB}
\sigma_{\rm RM} = 0.812 l_o \overline{ n_e} \sqrt {\langle B_{c,\parallel} \rangle ^2 (\frac{\Delta L}{l_o})^2 + B_r^2 (\frac{\langle L \rangle}{3l_o}) }
\end{equation}
where $\rm{\sigma_{RM}}$ $\sim$ 40 rad m$^{-2}$ is the weighted standard deviation in RM for the extragalactic sources that lie behind the SMC; $l_o$ $\sim$ 90 pc is the typical cell size along the line of sight, which we take to be similar to that in the LMC \citep{gaensler2005}; $\overline{n_e}$ = 0.039 cm$^{-3}$, is the mean electron density in the SMC as derived in \S~\ref{subsection:pulsardm}, $\langle B_{c,\parallel} \rangle$ $\approx$ $-$0.16 $\mu$G is the average SMC coherent field strength along the line of sight as obtained using ionized gas model 2; $\langle L \rangle$ $\approx$ 10 kpc is the average depth of the SMC along different sight lines; and $\Delta L$ $\approx$ 6 kpc is the standard deviation of the depth of the SMC between different sight lines (see \S~\ref{subsubsection:model2}). Using the above method, we find $B_r$ $=$ 19/$\sqrt{l_0}$ $\sim$  2 $\mu$G. Therefore, in the SMC, the random component of the magnetic field dominates over the coherent magnetic field along the line of sight.

A key prediction of our assumption that the RMs, optical starlight polarization and synchrotron intensity probe different projections of the same large-scale magnetic field is that the independently derived measurements of the random magnetic field must agree. Since the total synchrotron intensity probes the total magnetic field in the plane of the sky while the C-F method probes the ordered component in the plane of the sky, one can write: 
\begin{equation}
\langle B_{total,\perp}^2 \rangle = \langle B_{o,\perp}^2 \rangle  +\langle B_{r,\perp} ^2  \rangle 
\end{equation}
where $\langle B_{r,\perp} ^2  \rangle$ is the random magnetic field strength in the plane of the sky. If we assume that the random field is isotropic, then its strength is given by
\begin{equation}
B_r^2 =\frac{3}{2} \langle B_{r,\perp}^2 \rangle 
\end{equation}
Using $\langle B_{total,\perp} \rangle$ $\approx$ 3.2 $\mu$G (see \S~\ref{subsection:equipartition})  and $\langle B_{o,\perp} \rangle$ $\approx$ 1.6  $\mu$G (see \S~\ref{subsection:starlightcf}) leads to a random magnetic field strength of $\sim$ 3.4 $\mu$G. Since the estimate of the random field strength using the scatter of rotation measure agrees well with that obtained by combining the synchrotron intensity and starlight polarization measurements, our data demonstrate that our underlying assumptions are self-consistent.

\section{The 3D Magnetic Field Structure of the SMC}
\label{section:3Dfield}
We can combine the results of the RM study (\S~\ref{section:wimbfield}) and of the C-F method (\S~\ref{subsection:starlightcf}) to construct a 3D magnetic field vector for the SMC, assuming that the two methods probe the same field (in terms of strength, overall geometry and fluctuations).

The strength of the coherent magnetic field in the SMC is
\begin{equation}
B_{total,c} =  \sqrt{\langle B_{c,\parallel}\rangle^2+ \langle B_{c,\perp}^2 \rangle}=1.7\pm0.4 \mu G
 \end{equation}
where $ \langle B_{c,\parallel} \rangle$ = $-0.19$ $\pm$ 0.06 $\mu$G and $\langle B_{c,\perp} \rangle$ $\approx$ 1.6 $\pm$ 0.4~$\mu$G  denote the coherent fields found from Faraday rotation and optical starlight polarization, respectively. The three dimensional field is almost entirely in the plane of the sky. 

In order to more precisely determine the direction of the coherent magnetic field in the SMC, we need to transform into a cartesian coordinate system with the center of the SMC at the origin. We define our coordinate system such that the $x$-$y$ plane is the sky plane, the negative $x$ axis points towards the LMC's projection onto the sky plane, and the positive $z$ axis points along the vector joining the center of the SMC to the observer. In this coordinate system, the earth is located at (0, 0, 60) kpc  and the LMC is located at ($-$17, 0, 13) kpc. At the center of the SMC, the line-of-sight magnetic field is in the negative $z$ direction and has a strength of 0.19$\pm$ 0.06 $\mu$G, while the plane-of-the-sky magnetic field, with a magnitude of 1.6 $\pm$ 0.4 $\mu$G, makes an angle 4$^{\circ}$ (counterclockwise) with the positive $x$ axis as shown in \S~\ref{subsection:starlightcf}. Taking into account the ambiguity of the magnetic field direction in the plane of the sky, the coherent magnetic field vector in the SMC could be either
\begin{equation}
\label{eq:bvector1}
\vec{B_{c,1}}=1.6\hat{x}+0.1\hat{y}-0.19\hat{z}~~~~\rm{\mu G}
\end{equation}
or
\begin{equation}
\label{eq:bvector2}
\vec{B_{c,2}}=-1.6\hat{x}-0.1\hat{y}-0.19\hat{z}~~~\rm{\mu G}
\end{equation}

Equations~(\ref{eq:bvector1}) \& (\ref{eq:bvector2}) allow us to compute the possible angles that
the magnetic field vector makes with the characteristic axes of the
Magellanic System.  We consider two such axes: that defined by the path
from the LMC along the Magellanic Bridge to the SMC, and that defined
by the normal to the SMC disk.

Since the 3D structure of the Magellanic Bridge is not well known, we
here assume that the Bridge is parallel to $\vec{C}$, the vector which runs
from the center of the SMC to that of the LMC. $\vec{C}$ lies in the $x$-$z$
plane and has the form:
\begin{equation}
\vec{C}=17\hat{x}-13\hat{z}~~~\rm{kpc}
\end{equation}
This is a crude approximation, since the interaction between the
Magellanic Clouds most likely does not follow a straight line.

Separately, the normal to the plane of the SMC's disk is given by the
unit vector $\hat{n}$:
\begin{equation}
\label{eq:normal}
\hat{n}=-0.62\hat{x}-0.16\hat{y}-0.77\hat{z}
\end{equation}
Note that the angle between $\vec{C}$ and $\hat{n}$ is 92$^\circ$  (i.e., the SMC disk
is inclined by 2$^\circ$ degrees from the SMC-LMC axis).

We now consider the extent to which each of $\vec{B}_{c,1}$ and $\vec{B}_{c,2}$ are aligned
with $\vec{C}$ or are normal to $\hat{n}$. In the following discussion, we quote 90\%
confidence intervals in the statistical uncertainties in angles. We
consider any angle between vectors of less than $20^\circ$ to represent
broad alignment, and angles in the range $70^\circ - 110^\circ$ to
indicate rough perpendicularity (reflecting the additional systematic
uncertainties in our estimates of $\vec{C}$ and $\hat{n}$).

We find that the angle between $\vec{B}_{c,1}$ and $\vec{C}$ is 31$^{\circ {+8^\circ}}_{~-5^\circ}$. We have used Monte Carlo simulations with 50,000 random samplings to delineate the full probability distribution and find that the angle between $\vec{B}_{c,1}$ and $\vec{C}$ is consistent with alignment within $2.6\sigma$. On the other hand, $\vec{B}_{c,2}$ makes an angle 136$^{\circ {+4^\circ}}_{ ~-8^\circ}$ with $\vec{C}$. Monte Carlo simulations as described above show that any alignment between $\vec{B}_{c,2}$ and $\vec{C}$ is ruled out at $>$3.1$\sigma$.

Comparing the magnetic field vectors with $\hat{n}$, we find an angle between $\vec{B}_{c,1}$ and $\hat{n}$ of 123$^{\circ{+4^\circ}}_{~-8^\circ}$. We have used Monte Carlo simulations to find that the angle between $\vec{B}_{c,1}$ and $\hat{n}$ is consistent with $90^\circ$ at $\sim2.4\sigma$. The angle between $\vec{B}_{c,2}$ and $\hat{n}$ is  44$^{\circ{+7^\circ}}_{~-5^\circ}$. Monte Carlo simulations rule out any perpendicularity between  $\vec{B}_{c,2}$ and $\hat{n}$ at $4.2\sigma$.

The above calculations show that while at 90\% confidence level the SMC
magnetic field vector does not orient itself either with the Magellanic
Bridge or with the SMC disk, at a slightly higher confidence, the vector
$\vec{B}_{c,1}$ does indeed align with both the Bridge and the disk.  We thus
favor $\vec{B}_{c,1}$ as the more likely true magnetic field vector of the SMC
over $\vec{B}_{c,2}$. In this case, the possible alignment between $\vec{B}_{c,1}$ and $\vec{C}$ 
leaves open the Pan-Magellanic hypothesis  proposed by \cite{schmidt1970} and \cite{magalhaes1990}, i.e., that the SMC field orientation is
an imprint of the geometry of the overall Magellanic system.

To further test this Pan-Magellanic idea, additional RM studies of
extragalactic polarized sources behind the Magellanic Bridge will be
needed, to see whether the magnetic field in the Bridge potentially
also aligns with the vector $\vec{C}$. Meanwhile, the separate possibility that
$\vec{B}_{c,1}$ lies in the SMC disk (which as noted above, lies in a plane only
2$^\circ$ from the axis defined by the Bridge) provides an important
constraint on the origin of the magnetic field in the SMC, as we will
discuss fully in \S~\ref{section:discussion} below. We stress that the above analysis is
based on the assumption that the RMs and optical starlight polarization
probe the same large-scale field in the SMC.

\section{Discussion}
\label{section:discussion}
Our observations of the SMC demonstrate the existence of  a  large-scale coherent magnetic field. A coherent field cannot be explained by compression or stretching of a preexisting random field. The large scale dynamo is the usual mechanism invoked to produce a coherent magnetic field on galactic scales \citep{beck2000}. In this section, we explore which dynamo (or other) mechanisms might be responsible for producing this coherent field. 

\subsection{Ram Pressure Effects}
\label{subsection:rampressure}
When galaxies with large scale magnetic fields move rapidly through the intra-cluster medium (ICM), the field lines can be compressed, increasing the total magnetic energy of the system without dynamo action\citep{otmianowskamazur2003}. Therefore, it is reasonable to consider ram pressure as a mechanism that amplifies galactic magnetic fields. The maximum ram pressure considered in the 3D MHD model of \cite{otmianowskamazur2003} corresponds to a galaxy moving at a velocity of 1500 km s$^{-1} $ through an ICM of density 2 $\times$ 10$^{-3}$ cm$^{-3}$. The total magnetic energy is increased by a factor of $\sim$ 5 in their optimal model during the ram pressure event.  Simulated polarized intensity maps show characteristic features during different interaction phases with the ICM. Bright ridges are seen in the compressed region during the compression/stripping phase, while a large scale ``ring" field, resembling the field created by a dynamo mechanism, is seen during the gas re-accretion phase in the polarized intensity maps. No such features can be seen in single dish continuum data of the SMC  \citep{loiseau1987,haynes1991}. Furthermore, the space velocity of the galaxy used in the model of \cite{otmianowskamazur2003} is approximately three times larger than that of the velocity of the SMC with respect to the Galactic center \citep{kallivayalil2006}, while the density of the Milky Way halo is $\sim$ 10$^{-5}$ to 10$^{-4}$  cm$^{-3}$ at the distance of the SMC \citep{stanimirovic2002,sembach2006}. Therefore, the ram pressure effect on the SMC would be roughly 2 orders of magnitude weaker than for the simulations of   \cite{otmianowskamazur2003}. Also, it is unclear how ram pressure effects could generate a coherent large scale field from an initial field which might be incoherent. Therefore, we rule out the possibility of ram pressure effects generating the field in the SMC.

\subsection{The Mean-Field Dynamo}
\label{subsection:classical}

The $\alpha$-$\omega$ or mean-field dynamo requires turbulence to rise above or below the galactic disk to transform an azimuthal field into a poloidal one \citep{beck1996}. The radial component of the poloidal field is then transformed back into an azimuthal component by differential rotation of the disk. Although conservation of magnetic helicity can strongly suppress the $\alpha$ effect, it has been shown that this constraint on the mean field dynamo can be alleviated by flows between the disk and the halo, or by galactic outflows, which in turn allow the mean magnetic field to grow to a strength comparable to the equipartition value \cite[see for example][]{vishniac2004,shukurov2006}.

Dynamo action can be characterized by two parameters: $R_{\alpha}$ and $R_{\omega}$, given by \cite{ruzmaikin1988}

\begin{equation}
R_{\alpha}=\frac{3 l_o \Omega}{u_0}
\end{equation} 
\begin{equation}
R_{\omega}= \frac{3 s \frac {\partial \Omega}{\partial s} h_0 ^2 }{l_o u_0}
\end{equation} 
where $l_o$ is the outer scale of the turbulence, $s$ is the radial distance from the center of the galaxy, $h_0$ is the scale height of the gas disk and $\Omega$ is the angular velocity of the rotating disk. 
The typical speed, $u_0$, of turbulent motion of gas in the SMC can then be approximated by the velocity dispersion in HI, $u_0$ = 22 $\pm$ 2 km s$^{-1}$ \citep{stanimirovic2004}. 
It is generally believed that supernovae and superbubbles are the main drivers of turbulence in the Galactic disk \citep{mccray1979}, so $l_o$ is approximately the size of a supernova remnant or a superbubble. We assume that the ISM in the Milky Way, SMC and the LMC have comparable outer scales of turbulence, $l_o$ $\sim$ 90 pc \citep{gaensler2005} and gas disk scale heights $h_0$ $\sim$ 500 pc \cite[see for example][]{shukurov2004}. We use the SMC's HI rotation curve obtained by \cite{stanimirovic2004} to characterize its degree of differential rotation. Under the condition\footnotemark[6] \footnotetext[6]{For an energy injection scale of value $l_o$ $\sim$ 90pc, the condition that R$_{\omega}$ $\gg$ R$_{\alpha}$ is satisfied.} that R$_{\omega}$ $\gg$ R$_{\alpha}$, we can compute the dynamo number, $D$, a dimensionless parameter which determines the growth rate of the magnetic field \citep{ruzmaikin1988}:

\begin{equation}
\label{eq:dynamonumber}
D = R_\alpha R_\omega = \frac{ 9\Omega sh_0^2}  {u_{0}^2}  \frac {\partial \Omega}{\partial s} ~.
\end{equation} 
Note that the above equation is independent of the turbulent outer scale. Dynamo numbers at radii ranging from $s =$ 0.5 to 3.2 kpc are computed. In this range, the amount of shear in the disk of the SMC is given by $s \frac {\partial \Omega}{\partial s}$ $\sim$ 10$^{-16}$ s$^{-1}$, which is comparable to the shear in the Galactic disk near the sun\footnotemark[7] \footnotetext[7]{Adopting a value for Oort's constant A $\sim$ 15 km s$^{-1}$ kpc$^{-1}$}$\sim$ 5$\times$10$^{-16}$ s$^{-1}$. We obtain values of $|$D$|$ ranging from 0 to 4 in the SMC, while for the Milky Way, $|$D$|$ $\sim$ 20 in the solar vincinity \citep{shukurov2004}. The critical value for an exponential growth of the field is given by $|$$D_{\rm critical}$$|$ $\sim$ 8 $-$ 10, while a sub-critical dynamo number implies no growth \citep{shukurov2004}. We thus conclude that for the SMC, the classical mean field dynamo is not at work.

Using statistical studies of the SMC's neutral hydrogen, \cite{stanimirovic2001} found no characteristic scale of turbulence up to the size of the galaxy. This implies that the turbulent outer scale $l_o$ could be up to a few kpcs and the value R$_{\omega}$ would then be much smaller than R$_{\alpha}$. In this case, the dynamo number obtained using Equation~(\ref{eq:dynamonumber}) is no longer a good description of the field growth rate, since both the $\alpha$-$\omega$ and $\alpha^2$ dynamos (the latter is a dynamo driven by helical turbulence action alone) will operate. In this case the dynamo number $|$D$|$ would increase by $\sim$ 30$\%$ \citep{ruzmaikin1988}, which is not enough to rise the dynamo number above the critical level. Moreover, since the SMC experienced bursts of star formation triggered by tidal interactions $\sim$ 0.4 and 2.5 Gyrs ago \citep{zaritsky2004}, the additional energy injected into the ISM could have created outflows that would constantly disrupt the buildup of a large scale magnetic field produced by the $\alpha$-$\omega$ dynamo. We draw the conclusion that the mean-field dynamo is likely not responsible for the observed coherent field in the SMC.

\subsection{The Fluctuating Dynamo}
\label{subsection:fluctuatingdynamo}
It is thought that when the large scale dynamo is ineffective, as may occur in weakly rotating galaxies such as the SMC, the fluctuating dynamo (or the small-scale dynamo) can become important. The fluctuating dynamo, unlike the large-scale dynamo, can work without differential rotation in the galactic disk and can generate magnetic field with a correlation length similar to the energy carrying scale of the turbulence \citep{shukurov2004}. The fluctuating dynamo is believed to operate in small and slowly rotating galaxies with enhanced star formation, such as IC 10 \citep{chyzy2003}. The typical field amplification time scale is 10$^6$ to 10$^7$ years, much shorter than the standard dynamo growth rate. Signatures of random magnetic fields created by a fluctuating dynamo are isolated polarized non-themal regions coinciding with locations of star formation \citep{chyzy2003}. Since magnetic fields produced by a fluctuating dynamo are incoherent on galactic scales, they cannot be responsible for producing the observed coherent field in the SMC. However, the random field strength ($\sim$ 3$\mu$G) estimated in the SMC in \S~\ref{subsection:randomB} suggests that the fluctuating dynamo could be responsible for producing the random field component. Single dish radio continuum data of the SMC at multiple wavelengths show global diffuse synchrotron emission \citep{loiseau1987,haynes1991}, which also suggests that the random field strength in the SMC might be relatively high.

\subsection{The Cosmic-Ray Driven Dynamo}
\label{subsection:variations}

\cite{parker1992} proposed a cosmic-ray driven dynamo that has a much shorter amplification time scale than the standard mean-field dynamo. In this model, the driving force comes from cosmic rays injected into the galactic disk from the acceleration of charged particles in SNR shocks. Unlike the standard dynamo, this model incorporates a set of interacting forces including the buoyancy of cosmic rays, the Coriolis force, differential rotation and magnetic reconnection \citep{hanasz1998}. Differential rotation of the galactic disk is still required but the considerably larger $\alpha$ effect allows weakly rotating galaxies to achieve a supercritical dynamo number. The first numerical magneto-hydrodynamic (MHD) model of the CR-driven dynamo was developed by Hanasz \& Lesch (2004). They modeled a differentially rotating galaxy with a constant supply of cosmic rays and found that the large-scale magnetic field amplification time scale was about 250 Myrs. OB associations and frequent supernova explosions during the bursts of star formation in the SMC could result in a large cosmic ray flux, allowing the amplification of magnetic field in the SMC via the Parker-type dynamo. If the fast dynamo is responsible for the observed field due to the tidal triggered star formation episode $\sim$ 0.2 Gyrs ago, it would have just enough time to build up a galactic scale field before the tidal velocity field damps the dynamo effect \citep{kronberg1994,chyzy2004}. This can potentially also explain the coherent spiral field seen in the LMC \citep{gaensler2005}.

\cite{otmianowskamazur2000} modeled the magnetic field in the LMC-type irregular galaxy NGC 4449 using a value of $R_{\alpha}$ comparable to that of a fast dynamo. NGC 4449 is found to display ``fan" like structures that mimic magnetic spiral arms in polarized intensity. The Faraday rotation map of NGC 4449 suggests that the galaxy hosts a coherent field \citep{klein1996,chyzy2000}. \cite{otmianowskamazur2000} consider a model galaxy with a radius of $\sim$ 2.5 kpc and a maximum rotational velocity of about 30 km s$^{-1}$, which is similar to the SMC. No outflow from a bar and no random field were included. The value for  $l_0^2 \Omega /h_0 $ was 5 km s$^{-1}$ and the turbulent diffusivity ($\eta \sim l_0u_0/3$ ) was chosen to be 1.5$\times$10$^{26}$ cm$^2$ s$^{-1}$. These parameters are typical of a cosmic ray dynamo as shown by \cite{hanasz2004}. Evolving the modeled galaxy using the above parameters over $\sim$ 0.1 Gyr leads to an increase in the total magnetic energy. Also, this model is able to reproduce spiral-like field structure resembling the observation of NGC 4449. However, it does not include several possibly important physical processes. First,  the SMC is likely to be subjected to an injection of random field into the ISM due to a fluctuating dynamo (see \S~\ref{subsection:fluctuatingdynamo}), which this model does not account for. Second, the SMC has a rotation curve which peaks at $\sim$ 50 km s$^{-1}$ rather than the 30 km s$^{-1}$ used by  \cite{otmianowskamazur2000}. This results in a more effective $\omega$ effect, which increases the growth rate of the magnetic field, while random field injection increases the total magnetic energy of the galaxy faster. No simulated Faraday rotation map was produced by \cite{otmianowskamazur2000}, therefore, no direct comparison can be made between their model and our data. MHD models devoted to simulate the growth of the magnetic field in the SMC are needed in order to provide a definitive conclusion.

We have established above that it is possible for the cosmic-ray driven dynamo to produce the observed magnetic field in the SMC in terms of time scale arguments. Let us now consider whether this dynamo can explain the observed field geometry. In \S~\ref{section:3Dfield}, we showed that the 3D magnetic field vector of the SMC may lie in the disk of the galaxy and that it may align with the vector joining the Magellanic Clouds. A dynamo produces an azimuthal magnetic field that predominantly lies in
the disk of a galaxy  \cite[see][]{ruzmaikin1988}, and this could account
for the potential alignment of the SMC's magnetic field with the SMC
disk as calculated in \S~\ref{section:3Dfield}. If the field is aligned with the Bridge
rather than the disk (as also allowed by the range of angles calculated
in \S~\ref{section:3Dfield}), this could be understood as resulting from ongoing tidal
interactions between the Magellanic Clouds, which could provide a slight
realignment of the overall field orientation.

If the cosmic-ray driven dynamo is the underlying mechanism that produces the magnetic field in the SMC, it also needs to explain the unidirectional field lines seen across the galaxy. The magnetic field configuration in a galaxy can be decomposed into different dynamo modes \citep{beck1996}. The strongest dynamo mode in an axisymmetric disk is the $m=0$ mode, followed by a weaker bisymmetric ($m=1$) mode. It has been suggested by \cite{moss1995} that tidal interactions can generate bisymmetric magnetic fields in galaxies, provided that the axisymmetric mode is already at work. Observations show that in interacting galaxies, such as M51 and M81, the bisymmetric mode can be important \citep{krause1989}. According to the 3D mean field dynamo model studied by \cite{vogler2001}, non-axisymmetric gas motion is induced in galactic disks during tidal interaction, and can damp the usual dominant $m=0$ mode and excite the $m=1$ mode when the induced tidal velocity is small. An axisymmetric magnetic field would exhibit a change in the sign of RM across the disk of the galaxy when viewed edge on whereas a bisymmetric magnetic field would vary double-periodically with the azimuthal angle \citep{krause1989}. It is unclear how the superposition of a $m=0$ mode and a $m=1$ mode could produce unidirectional field lines with negative RM across the SMC, because a superposition of higher order modes will result in more RM sign changes across the galaxy disk when viewed edge on. 

The observed unidirectional magnetic field lines and the possible alignment of the field with the Magellanic Bridge could be explained as follows. Cosmic ray driven dynamo produces a predominately azimuthal magnetic field in the SMC disk; this field is then stretched tidally along the SMC-LMC axis, maintaining its orientation when projected onto the plane of the sky (to produce starlight polarization vectors of similar orientation). Note that RM is non-zero only when the average line of sight electron density is non-zero. It is possible that the field lines in the SMC do close, that is, there are sight lines along which field lines do point towards us, but only at locations with low ${\rm EM}$ off the main body of the SMC. Only half of the displaced magnetic loop, whose line of sight component is directed away from us, would then be observed. The other half of the loop whose line of sight component is directed towards us would not show positive RMs, as it should coincide with regions of low ${\rm EM}$ . 

To summarize, the cosmic-ray driven dynamo is a possible field generation mechanism for the SMC but has difficulties explaining the observed magnetic field geometry. One has to explain the fact that the observed field is unidirectional and that it potentially lies in the disk of the SMC and aligns with the Magellanic Bridge. Current observational data are not sufficient to rule out/prove the cosmic-ray driven dynamo; further observational tests are needed.

\section{Conclusions}
\label{section:conclusions}
We have measured the Faraday rotation of extragalactic polarized sources behind the Small Magellanic Cloud to determine the SMC's magnetic field strength and geometry. Our study reveals that the SMC has a galactic-scale field of  0.19 $\pm$ 0.06 $\mu$G directed coherently away from us along the line of sight. Optical polarization data on stars in the SMC are re-analyzed using the Chandradsekhar-Fermi method to give an ordered component of the magnetic field in the plane of the sky, of strength 1.6 $\pm$ 0.4 $\mu$G. Under the assumption that the Faraday rotation measures and optical starlight polarization probe the same underlying large scale field in the SMC, we have constructed a 3D magnetic field vector of the SMC. It is found that this magnetic field vector possibly aligns with the Magellanic Bridge. This potential alignment needs to be verified by future studies of RMs towards extragalactic sources behind the Magellanic Bridge. The random magnetic field strength in the SMC derived from RM data alone and that derived by combining the results of the C-F method with equipartition were found to be in agreement ($\sim$ 3 $\mu$G). This implies that our underlying assumption, that these 3 independent methods probe different components of the same large scale field, is self-consistent.

The SMC is a slowly rotating galaxy, for which the standard mean-field dynamo is not expected to be at work because of the subcritical dynamo number. The cosmic-ray driven dynamo has a short enough amplification time scale to explain the observed coherent field. With modifications by tidal interactions, the field generated by the cosmic-ray driven dynamo could potentially be aligned with the Magellanic Bridge. However, this model faces difficulties in explaining the observed uni-directional field lines. Therefore, the origin of the magnetic field in the SMC is still an open question which needs to be followed up with more observations. The relatively small number of background rotation measures makes it difficult to interpret the observed RMs in detail. Future observations of the SMC with the Square Kilometre Array will provide  $\sim$ 10$^5$ RMs in a field of 40 square degrees surrounding the SMC \citep{beck2004b}; with which different possible origins of the magnetic field in the SMC can be fully evaluated \citep{stepanov2008}.

\textbf{Acknowledgements}
We thank Joseph Gelfand for carrying out the ATCA observations, Anvar Shukurov, Erik Muller, Alyssa Goodman and Douglas Finkbeiner for useful discussions, and Rainer Beck, Marita Krause and Ellie Berkhuijsen for their help and hospitality during S.~A.~M. 's visit to the Max Planck Institute for Radio Astronomy. M.~H. acknowledges support from the National Radio Astronomy Observatory (NRAO), which is operated by Associated Universities Inc., under cooperative agreement with the National Science Foundation. This research was supported by the National Science Foundation through grant AST-0307358 to Harvard College Observatory. The Australia Telescope Compact Array is part of the Australian Telescope, which is funded by the Commonwealth of Australia for operation as a National Facility managed by CSIRO. The Southern H-Alpha Sky Survey Atlas (SHASSA) is supported by the NSF. 


{\it Facilities:} ATCA

\begin{appendix}
\section{Extinction Correction of  H$\alpha$ Emission}
\label{appendix:hacorrection}
We here describe the procedure to derive the intrinsic H$\alpha$ intensity of the SMC. We assume that dust is well mixed with H$\alpha$ emitting gas for both the SMC and the Milky Way. The observed H$\alpha$ emission is given by
\begin{equation}
I_{\rm observed} =  \frac {I_{\rm instrinic,SMC}}{\tau_{\rm H\alpha,SMC}} (1-e^{\tau_{\rm H\alpha,SMC}}) e^{-\tau_{\rm H\alpha,MW}} + \frac{I_{\rm intrinsic,MW}} {\tau_{\rm H\alpha,MW}} (1-e^{-\tau_{\rm H\alpha,MW}})
\end{equation}
where $\tau_{\rm H\alpha,SMC}$ is the optical depth of  H$\alpha$ in the SMC, $\tau_{\rm H\alpha,MW}$ is the optical depth of H$\alpha$ in the Milky Way, $I_{\rm intrinsic,MW}$ is the intrinsic H$\alpha$ emission of the Milky Way, and $I_{\rm intrinsic,SMC}$ is the intrinsic H$\alpha$ emission of the SMC. The second term in the above equation can be estimated by the observed off source H$\alpha$ intensity in the regions surrounding the SMC.

The uncertainty in estimating the intrinsic H$\alpha$ intensity mainly results from the location of the dust with respect to the H$\alpha$ emitting regions. The upper estimate of $I_{\rm instrinic,SMC}$ can be found by placing all the dust behind the H$\alpha$ emitting region in the SMC, so that what we observe is the intrinsic H$\alpha$ intensity extincted only by the foreground Milky Way dust. The lower estimate of $I_{\rm instrinic,SMC}$ can be found by placing all the dust in front of the H$\alpha$ emitting region in the SMC.

\section{A Model to estimate the Random Magnetic Field Strength}
\label{appendix:randomB}

We construct this model based on \cite{gaensler2001,gaensler2005} and ionized gas model 2 (see \S~\ref{subsubsection:model2}), for which case we assume that the average electron density ($\overline{n_e}$) along different lines of sight is the same but the depth of the SMC varies from one sight line to the other. From model 3, the mean depth through the SMC is $\langle L \rangle$ $\approx$ 10 kpc with a standard deviation $\Delta L$ $\approx$ 6 kpc. Suppose that the depth of the SMC through a particular sight line is $L$, divided up into cells of linear size $l_o$. The total number of cells one looks through along the line of sight is given by

\begin{equation}
N= \frac {L}{l_o}
\end{equation}

Within each cell, we suppose that the magnetic field is composed of a coherent component of strength $B_{c}$ (same direction and strength from cell to cell), whose strength along the line of sight is $\langle B_{c,\parallel} \rangle$ $\approx$ 0.16$\mu$G, and a random component of strength $B_r$ oriented at an angle $\theta_{\rm cell,i}$ with respect to the line of sight. The component of the random field along the line of sight is 
\begin{equation}
B_{r,\parallel} = B_r {cos} \theta_{\rm cell,i}
\end{equation}
The line of sight magnetic field strength in a cell is given by
\begin{equation}
{B_\parallel} = \langle B_{c,\parallel} \rangle + B_{r,\parallel} =  \langle B_{c,\parallel} \rangle  + B_r {cos}\theta_{\rm cell,i}
\end{equation}
In addition, we assume that the random component is coherent within each cell but that $\theta$ varies randomly from cell to cell. Different levels of Faraday rotation will be experienced by the incident light rays because they pass through different series of cells and different numbers of cells. Linearly polarized light which passes through a single cell in the SMC experiences a Faraday rotation given by
\begin{equation}
{\rm RM}_{1-{\rm cell}} = 0.812 n_{e_{\rm cell,i}} l_o \overline{B_{\parallel}} =  0.812 n_{e_{\rm cell,i}} l_o ( \langle B_{c,\parallel} \rangle  + B_r cos\theta_{\rm cell,i})
\end{equation}
After passing through $N$ cells, the incident radiation experiences a Faraday rotation of
\begin{equation}
{\rm RM}_{N-{\rm cells}} = 0.812 l_o B_r \sum_{i=1}^N  n_{e_{\rm cell,i}}  cos\theta_{\rm cell,i} + 0.812 l_o  \langle B_{c,\parallel} \rangle  \sum_{i=1}^N  n_{e_{\rm cell,i}} 
\end{equation}
where 
\begin{equation}
\sum_{i=1}^N  n_{e_{\rm cell,i}} = \overline{{n_e}} N
\end{equation}
Since the electron density does not correlate with the orientation of the random field  in individual cells, 
\begin{equation}
\sum_{i=1}^N  n_{e_{\rm cell,i}} cos\theta_{\rm cell,i}  = \overline{{n_e}} \sum_{i=1}^N  cos\theta_{\rm cell,i} ~.
\end{equation}
One can rewrite the expression for the rotation measure of the radiation after passing through N cells as
\begin{equation}
{\rm {RM}_{N-{\rm cells}}} = 0.812 l_o B_r \overline{{n_e}}  \sum_{i=1}^N  cos\theta_{\rm cell,i} + 0.812 l_o  \langle B_{c,\parallel} \rangle  N \overline{{n_e}} 
\end{equation}
Averaging across different sight lines, the mean RM through the SMC is given by
\begin{equation}
\langle {\rm RM} \rangle  = 0.812 l_o  \langle B_{c,\parallel} \rangle  \langle N \rangle 
\end{equation}
where $\langle N \rangle = \langle L \rangle /l_o$ is the average number of cells along different sight lines.

Using the central limit theorem for large $N$, the standard deviation of RM through the SMC can be expressed as
\begin{equation}
\sigma_{\rm RM} = 0.812 l_o \overline{{n_e}} \sqrt { {\langle B_{c,\parallel} \rangle} ^2 (\frac{\Delta L}{l_o})^2 + {B_r}^2 (\frac{\langle L \rangle}{3l_o}) } ~.
\end{equation}

\end{appendix}

\clearpage
\LongTables
\begin{landscape}
\begin{deluxetable}{lll}
\tabletypesize{\footnotesize}
\tablecaption{A list of symbols used in this paper}
\tablewidth{0pc} 
\tablehead{
\colhead{Symbol} & \colhead{Physical Quantity} & \colhead{Section in which first used}}
\startdata
$\alpha$ & spectral index of synchrotron emission defined by $I_\nu\propto\nu^{-\alpha}$  & \S~\ref{subsection:previousstudies}\\
a & angular offset in degrees eastward from RA = 0 &\S~\ref{subsection:radiodata}\\
$B_{\parallel}(l)$ & magnetic field as a function of path length $l$  along the line of sight in units of $\mu$G & \S~\ref{subsection:faradayrotation}\\
$\overline{B_\parallel}$ & average magnetic field strength along the line of sight in $\mu$G & \S~\ref{subsection:ionizedgasmodels} \\
$ \langle B_{o,\perp} \rangle $ & average magnetic field strength in the plane of the sky in units of $\mu$G & \S~\ref {subsection:osp}\\
$B_{\parallel,i}$ & individual measurement of line-of-sight magnetic field $\overline{B_\parallel}$ through the SMC in units of $\mu$G  & \S~\ref{subsubsection:modelsummary} \\
$\langle B_{c,\perp} \rangle $ & strength of the coherent component of the magnetic field in the plane of the sky in units of $\mu$G  & \S~\ref{subsection:starlightcf} \\
$ \langle B_{total,\perp} \rangle $ & strength of the total equipartition magnetic field in the plane of the sky in units of $\mu$G & \S~\ref{subsection:equipartition} \\
$B_r$ & strength of random magnetic field in units of $\mu$G  & \S~\ref{subsection:randomB}\\
$\sigma_{B,i}$ & uncertainty associated with an individual magnetic field measurement $B_{\parallel,i}$ in units of $\mu$G & \S~\ref{subsubsection:modelsummary} \\
$B_{r,\parallel}$ & strength of random magnetic field parallel to the line of sight in units of $\mu$G & \S~\ref{appendix:randomB} \\
$\langle B_{c,\parallel} \rangle $ & coherent magnetic field parallel to the line of sight averaged across the SMC in units of $\mu$G & \S~\ref{subsubsection:modelsummary} \\
$\langle B_{r,\perp} \rangle $ & random magnetic field strength perpendicular to the line of sight in units of $\mu$G  & \S~\ref{subsection:randomB}\\
$ B_{total,c} $ & total coherent magnetic field strength in units of $\mu$G &  \S~\ref{section:3Dfield}\\
$\vec{B}_{c,1}$, $\vec{B}_{c,2}$ & the 3D magnetic field vector of the SMC, subscripts 1 and 2 indicates the two vectors whose &\\
& line-of-sight component is the same but have oppositely directed plane-of-the-sky component & \S~\ref{section:3Dfield}\\
$\gamma_H$ & equivalent molecular weight of the ISM for SMC abundance  & \S~\ref{subsection:starlightcf} \\
$\gamma_e$ & power law index of electron energy distribution in cosmic rays& \S~\ref{subsection:equipartition} \\
$\vec{C}$ & the vector joining the center of the LMC to the center of the SMC  &  \S~\ref{section:3Dfield}\\
$\Delta\phi $ & angle that the polarization plane rotates through in radians & \S~\ref{subsection:faradayrotation}\\
$\delta_p$ & angular deviation of $\theta_p$ from $\langle \theta_p \rangle $  & \S~\ref{subsection:osp} \\
$D$ & dynamo number  & \S~\ref{subsection:classical}\\
$D_{\rm critical}$ & critical dynamo number  & \S~\ref{subsection:classical}\\
DM & dispersion measure in units of pc cm$^{-3}$ & \S~\ref{subsection:pulsardm} \\
$\rm{\langle DM_{SMC,pulsar} \rangle} $ & average dispersion measure towards SMC pulsars, after foreground subtraction & \S~\ref{subsection:pulsardm} \\
$\rm{\langle DM_{SMC} \rangle}$ & average dispersion measure through the SMC &  \S~\ref{subsection:pulsardm} \\
$\rm{\sigma_{DM}}$ & standard deviation of pulsar DMs in the SMC & \S~\ref{subsection:pulsardm}\\
EM & emission measure in units of pc cm$^{-6}$ &  \S~\ref{subsection:emissionmeasure} \\
$\rm{\langle EM_{SMC} \rangle} $ & average emission measure through the SMC in pc cm$^{-6}$ &  \S~\ref{subsubsection:model2} \\
$\rm{EM_{source}}$ & emission measure towards an individual extragalactic source behind the SMC  & \S~\ref{subsubsection:model2} \\
$f$ & filling factor of thermal electrons along the line of sight &  \S~\ref{subsubsection:model2} \\
$fL$ & occupation length in units of pc of thermal electrons along the line of sight & \S~\ref{subsubsection:model3} \\
$h_0$ & scale height of galactic disk in pc & \S~\ref{subsection:classical}\\
$\eta$ & turbulent diffusivity in units of cm$^2$ s$^{-1}$ & \S~\ref{subsection:variations} \\
$\theta_{cell,i}$  & orientation of the random magnetic field with respect to the line of sight in cell $i$ & \S~\ref{appendix:randomB} \\
$I_\nu$ & specific intensity of synchrotron emission in erg s$^{-1}$ cm$^{-2}$ Hz$^{-1}$ sr$^{-1}$ &  \S~\ref{subsection:previousstudies}\\
$I_{\rm H\alpha,intrinsic,SMC}$ & intrinsic H$\alpha$ intensity of the SMC in units of Rayleighs & \S~\ref{subsection:emissionmeasure} \\
$I_{\rm observed}$ & observed H$\alpha$ intensity towards the SMC & \S~\ref{appendix:hacorrection} \\
$I_{\rm instrinic,SMC} $ & H$\alpha$ intensity of the SMC after extinction correction for both the Milky Way and the SMC & \S~\ref{appendix:hacorrection} \\
$I_{\rm intrinsic,MW}$ & H$\alpha$ intensity of the Milky Way after extinction correction & \S~\ref{appendix:hacorrection} \\ 
$k$ & dimensionless galactic dust-to-gas ratio defined in Equation 9 & \S~\ref{subsection:emissionmeasure} \\
$K$ & energy density of cosmic rays in the galaxy in units of erg$^{-1}$ cm$^{-3}$ & \S~\ref{subsection:equipartition} \\
$K_0$ & ratio of number densities of protons to electrons in cosmic rays accelerated in SNRs & \S~\ref{subsection:equipartition} \\
$dl$ & differential path length along the line of sight in units of pc& \S~\ref{subsection:faradayrotation}\\
$l$ & path length along the line of sight in units of pc & \S~\ref{subsection:faradayrotation} \\
$L$ & total path length along the line of sight & \S~\ref{subsection:pulsardm} \\
$\langle L_{\rm SMC} \rangle$ & average depth of the SMC in pc & \S~\ref{subsubsection:model2} \\
$L_{\rm source}$ & path length through the SMC to an extragalactic source in pc  & \S~\ref{subsubsection:model2} \\
$L_{\rm synchrotron}$ & depth of the synchrotron emitting layer in the SMC in cm & \S~\ref{subsection:equipartition} \\
$\langle L \rangle $ & average path length through the SMC to extragalactic sources in units of pc & \S~\ref{subsubsection:model2}\\
$\Delta L $ & standard deviation of $L_{\rm source}$ through different sight lines in the SMC & \S~\ref{subsubsection:model2}\\
$l_o$ & Turbulence outer scale/ typical cell size in the SMC in units of pc  & \S~\ref{subsection:randomB},  \S~\ref{subsection:classical}\\
$\lambda $ & wavelength in meters & \S~\ref{subsection:faradayrotation}\\
$m$ & dynamo mode & \S~\ref{subsection:variations} \\
$m_H$ & mass of hydrogen atom &  \S~\ref{subsection:starlightcf} \\
$\xi(\lambda_{\rm H\alpha})$ & extinction of the Milky Way and of the SMC at the wavelength of H$\alpha$ & \S~\ref{subsection:emissionmeasure} \\
$\nu$ & frequency in Hz &  \S~\ref{subsection:previousstudies} \\
$\hat{n}$ & unit vector normal to the SMC's HI disk & \S~\ref{section:3Dfield} \\
n & number of sight lines through the SMC & \S~\ref{subsubsection:modelsummary} \\
$n_e(l)$ & electron density distribution as a function of path length $l$  in units of cm$^{-3}$ & \S~\ref{subsection:faradayrotation}\\
$\overline{n_e}$ & average electron density along the line of sight & \S~\ref{subsection:pulsardm}\\
$\langle{n_e}\rangle$ & mean electron density in the SMC in units of cm$^{-3}$ & \S~\ref{subsection:pulsardm}\\
$\overline{n_e^2}$ & average of the square of the electron density along the line of sight & \S~\ref{subsection:emissionmeasure} \\
$n_{\rm cloud}$ & electron density in an ionized gas cloud & \S~\ref{subsubsection:model2} \\
$\overline{n_{\rm cloud}}$ & average electron density in ionized gas clouds along the line-of-sight & \S~\ref{subsubsection:model2} \\
$n_{e_{\rm cell,i}}$ & electron density in cell $i$ along the line of sight & \S~\ref{appendix:randomB} \\
$n_H$ & number density of hydrogen in the ISM of the SMC &  \S~\ref{subsection:starlightcf} \\
$N_{\rm HI} $ & neutral hydrogen column density in units of cm$^2$& \S~\ref{subsection:emissionmeasure} \\
 $N$ & number of cells along a line of sight & \S~\ref{appendix:randomB} \\
 $\langle N \rangle$ & average number of cells when looking through the SMC along the line of sight & \S~\ref{appendix:randomB} \\
$P_i$ & weights =$1/\sigma_{B,i}^2$ & \S~\ref{subsubsection:modelsummary} \\
$\overline{P}$ & weighted mean of the weights $P_i$ & \S~\ref{subsubsection:modelsummary} \\
Q & quality of fit of the least square fit of $\Delta \phi $ against $\lambda^2$ & \S~\ref{subsection:radiodata}\\
RM & rotation measure in units of rad m$^{-2} $& \S~\ref{subsection:faradayrotation}\\
$\rm{RM_{observed}}$ & observed RM of an extragalactic source & \S~\ref{subsection:faradayrotation}\\
$\rm{RM_{SMC}}$ & rotation measure through the SMC & \S~\ref{subsection:faradayrotation}\\
$\rm{RM_{Intrinsic}}$ & intrinsic rotation measure of an extragalactic source & \S~\ref{subsection:faradayrotation}\\
$\rm{RM_{IGM}}$ & rotation measure originating from the intergalactic medium & \S~\ref{subsection:faradayrotation}\\
$\rm{RM_{Milky Way}}$ & rotation measure due to the Milky Way foreground & \S~\ref{subsection:faradayrotation}\\
$\langle {\rm RM} \rangle$ & weighted mean of the RM of extragalactic sources through the SMC & \S~\ref{appendix:randomB} \\
${\rm RM}_{1-{\rm cell}}$ & rotation measure through 1 cell & \S~\ref{appendix:randomB} \\
${\rm RM}_{N-{\rm cells}}$ &  rotation measure through N cells & \S~\ref{appendix:randomB} \\
$R_\alpha$ & parameter that characterizes the $\alpha$-effect & \S~\ref{subsection:classical}\\
$R_\omega$ & parameter that characterizes the $\omega$-effect  & \S~\ref{subsection:classical}\\
$\rho$ & mass density of the phase of the ISM containing dust in g cm$^{-3}$ & \S~\ref{subsection:osp}\\
$\sigma_{\rm RM}$ & weighted standard deviation of RMs & \S~\ref{subsection:randomB}\\
$\rm{\sigma_{spatial,1D}}$ & one dimensional spatial dispersion of the radio pulsars in the SMC & \S~\ref{subsection:pulsardm}\\
$\sigma_{v_{los}}^2$ & line of sight velocity dispersion of the phase of the ISM containing dust & \S~\ref{subsection:osp} \\
$SEM_w$ & weighted standard error in the mean line-of-sight magnetic field & \S~\ref{subsubsection:modelsummary} \\
$s$ & radial distance from the center of the galaxy & \S~\ref{subsection:classical}\\
$T_e$ & electron temperature of the WIM in Kelvins  & \S~\ref{subsection:emissionmeasure} \\
$\tau_{\rm H\alpha,SMC}$ & H$\alpha$ optical depth of the SMC & \S~\ref{appendix:hacorrection} \\
$\tau_{\rm H\alpha,MW}$ & H$\alpha$ optical depth of the Milky Way & \S~\ref{appendix:hacorrection} \\
$\tau_{\rm B}$ & extinction optical depth in the optical B band & \S~\ref{subsection:emissionmeasure} \\
$\tau_{\rm H\alpha}$ & optical depth of the SMC/Milky Way in the H$\alpha$ line& \S~\ref{subsection:emissionmeasure} \\
$\theta_p$ & position angle of starlight polarization in radians  & \S~\ref{subsection:osp} \\
$\langle \theta_p \rangle $ & weighted mean of $\theta_p$ towards stars in the SMC  & \S~\ref{subsection:osp}\\
$u_0$ & turbulent velocity in the galaxy in cm s$^{-1}$  & \S~\ref{subsection:classical}\\
$\Omega$ & angular velocity of galactic rotation  & \S~\ref{subsection:classical}\\
\enddata
\label{table:glossary}
\end{deluxetable}
\clearpage
\end{landscape}

\begin{deluxetable}{ll}
\tablecaption{Properties of the SMC}
\tablewidth{0pc}
\tablehead{
\colhead{Parameter } &  \colhead{   }
}
\startdata
Right ascension (J2000) \tablenotemark{a} & $01^{h}05^{m} $  \tablenotemark{b}\\
Declination (J2000) \tablenotemark{a} &  $-72^{d}49.7^{m}$  \tablenotemark{b}\\ 
Galactic coordinates(\rm{l},b) & (302.8,-44.6)  \tablenotemark{c}\\
Galacto-centric coordinate (X,Y,Z) (kpc) & (15.3, -36.9,-43.3 )  \tablenotemark{c}\\
Galacto-centric Space Velocity ($v_X$,$v_Y$,$v_Z$) (km s$^{-1}$) & (-87$\pm$48, -247$\pm$42, 149$\pm$37)  \tablenotemark{c}\\
Diameter on Sky & $\approx$7 $^{\circ}$  \tablenotemark{d}\\
Distance (kpc) & $\approx$ 60 \\
Inclination of HI disk & 40$^{\circ}$ $\pm$ 20$^{\circ}$ \tablenotemark{b} \\
Kinematic PA of major axis \tablenotemark{e} & $\sim$ 40$^{\circ}$ \tablenotemark{b}\\
HI Mass & 4.2$\times$ 10$^8$ M$_{\odot}$ \tablenotemark{b}\\
\enddata
\label{table:smcproperties}
\tablenotetext{a}{The apparent kinematic center of the SMC}
\tablenotetext{b}{\cite{stanimirovic2004}}
\tablenotetext{c}{\cite{kallivayalil2006}}
\tablenotetext{d}{\cite{westerlund1990}}
\tablenotetext{e}{The position angle (PA) is measured from north through east}
\end{deluxetable}
\clearpage

\clearpage
\LongTables
\begin{deluxetable}{lllr}
\tablecolumns{5} 
\tablewidth{100pt} 
\tablecaption{\scriptsize Rotation Measures of Extragalactic Sources in the Observed Field} 
\vspace{0.01in}
\tablehead{   
\colhead{{Source Name}\tablenotemark{a}} &    \colhead{RA(J2000)(hms)} &     \colhead{DEC(J2000)(dms)}   &     \colhead{RM(rad m$^{-2}$)}} 
\label{table:rawrms}
\startdata 
005 & 1:41:55.85 & $-$69:41:39.94 & $-$12 $\pm$ 18 \\
006 & 1:41:28.51 & $-$70:16:35.80 & $+$ 2 $\pm$ 24 \\
007 & 1:41:16.35 & $-$70:15:04.00 & $+$ 40 $\pm$ 25 \\
008 & 1:33:48.46 & $-$69:28:43.82 & $+$ 3 $\pm$ 14 \\
009 & 1:33:47.71 & $-$69:28:38.37 & $+$ 2 $\pm$ 14\\
010 & 1:33:39.56 & $-$69:28:37.32 & $-$16 $\pm$ 20 \\
011 & 1:28:45.87 & $-$69:36:16.12 & $-$5 $\pm$ 15 \\
012 & 1:29:44.77 & $-$70:35:31.04 & $+$18 $\pm$ 24 \\
013 & 1:21:49.98 & $-$69:56:43.37 & $+$21 $\pm$ 9 \\
014 & 1:21:44.40 & $-$69:57:21.11 & $+$28 $\pm$ 19 \\
015 & 1:28:16.99 & $-$75:12:58.75 & $+$35 $\pm$ 13 \\
016 & 1:28:08.78 & $-$75:12:51.83 & $+$42 $\pm$ 14 \\
017 & 1:22:57.30 & $-$75:15:04.66 & $+$30 $\pm$ 11 \\
018\tablenotemark{*} & 1:10:35.37 & $-$72:28:07.70 & $-$82 $\pm$ 27 \\
019\tablenotemark{*} & 1:10:55.99 & $-$73:14:11.49 & $-$37 $\pm$ 14 \\
020\tablenotemark{*} & 1:10:50.67 & $-$73:14:25.42 & $+$3 $\pm$ 13 \\
021\tablenotemark{*} & 1:10:48.65 & $-$73:14:29.14 & $+$6$\pm$ 5 \\
022 & 1:01:32.49 & $-$69:39:14.21 & $-$8 $\pm$ 32 \\
023 & 1:03:33.17 & $-$75:06:57.46 & $+$6 $\pm$ 33\\
024 & 0:57:15.86 & $-$70:40:47.01 & $+$44 $\pm$ 24 \\
025\tablenotemark{*} & 0:56:45.05 & $-$72:51:59.06 & $-$30 $\pm$ 41 \\
026 & 0:52:23.71 & $-$75:25:48.72 & $-$5 $\pm$ 17 \\
027 & 0:47:40.77 & $-$75:30:11.36 & $+$21 $\pm$ 4 \\
028 & 0:42:38.74 & $-$70:01:34.56 & $+$34 $\pm$ 24 \\
029 & 0:40:47.68 & $-$71:45:59.63 & $+$18 $\pm$ 26 \\
030 & 0:39:39.59 & $-$71:41:42.44 & $+$41 $\pm$ 21 \\
031 & 0:38:01.40 & $-$72:52:10.65 & $+$21 $\pm$ 24 \\
032 & 0:37:54.79 & $-$72:51:56.66 & $+$1 $\pm$ 13 \\
033 & 0:34:28.44 & $-$73:35:26.77 & $+$12 $\pm$ 24 \\
034 & 0:34:25.83 & $-$73:35:13.86 & $+$10 $\pm$ 27 \\
035 & 0:34:14.98 & $-$73:33:28.08 & $+$43 $\pm$ 18 \\
036 & 0:34:15.58 & $-$73:33:34.29 & $+$53 $\pm$ 14 \\
037 & 0:34:15.20 & $-$73:33:16.16 & $+$42 $\pm$ 22 \\
038 & 0:29:20.10 & $-$75:40:08.70 & $+$7 $\pm$ 48 \\
039 & 0:32:31.35 & $-$73:06:50.38 & $+$50 $\pm$ 30 \\
040 & 0:32:30.92 & $-$69:24:28.54 & $+$9 $\pm$ 48 \\
041 & 0:26:08.35 & $-$73:23:15.03 & $+$51 $\pm$ 20 \\
042 & 0:24:11.37 & $-$73:57:15.95 & $+$54 $\pm$ 8 \\
043 & 0:23:37.31 & $-$73:55:30.19 & $+$51 $\pm$ 8 \\
044 & 0:22:21.97 & $-$74:28:21.27 & $+$40 $\pm$ 9 \\
045 & 0:29:28.04 & $-$69:34:35.04 & $+$24 $\pm$ 12 \\
046 & 0:29:26.11 & $-$69:34:46.23 & $+$26 $\pm$ 17 \\
047 & 0:22:15.40 & $-$74:28:14.62 & $+$ 63 $\pm$ 19 \\
048 & 0:22:02.76 & $-$74:27:22.53 & $+$  40 $\pm$ 23 \\
082 & 0:14:24.12 & $-$73:39:05.15 & $+$  29 $\pm$ 33 \\
087 & 0:11:56.37 & $-$73:49:56.30 & $+$ 84 $\pm$ 41 \\
115 & 0:31:36.77 & $-$70:33:18.49 &  $+$ 67 $\pm$ 70 \\
116 & 0:28:41.87 & $-$70:45:19.61 & $+$ 60 $\pm$ 38 \\
117 & 1:39:48.32 & $-$69:33:27.44 & $+$ 16 $\pm$ 43 \\
118 & 1:18:09.61 & $-$69:46:04.36 & $+$ 22 $\pm$ 51 \\
119 & 1:18:16.62 & $-$69:51:50.35 & $+$ 20 $\pm$ 52\\
120 & 1:22:45.57 & $-$69:44:18.52 & $+$ 31 $\pm$ 38 \\
121\tablenotemark{*} & 1:10:45.44 & $-$72:28:52.19 & $-$69 $\pm$ 44 \\
122 & 1:19:17.56 & $-$71:05:42.40 &  $+$ 43 $\pm$ 52 \\
123 & 0:34:01.18 & $-$70:26:29.06 & $+$ 41 $\pm$ 46 \\
124 & 0:21:38.21 & $-$69:26:16.61 & $+$ 12 $\pm$ 47 \\
125 & 0:28:36.06 & $-$69:33:40.34 & $+$ 16 $\pm$ 34 \\
126 & 0:46:39.63 & $-$69:57:12.48 & $+$ 20 $\pm$ 32\\
127 & 0:42:24.66 & $-$70:02:47.47 & $-$5 $\pm$ 30 \\
128\tablenotemark{*} & 0:49:34.99 & $-$72:19:03.34 & $-$204 $\pm$ 76 \\
130 & 0:26:06.34& $-$73:23:10.97 & $+$ 46 $\pm$ 27 \\
131 &  1:37:02.32 & $-$73:04:17.75 & $-$16 $\pm$ 32 \\
132 &  1:37:05.69 & $-$73:04:14.74 & $+$ 4 $\pm$ 51 \\
133 &  1:46:39.13 & $-$72:48:55.80 & $-$8 $\pm$ 49 \\
134\tablenotemark{*} &  1:29:30.54 & $-$73:33:12.41 & $-$385 $\pm$ 56 \\
135\tablenotemark{*} & 1:33:30.53 & $-$73:03:06.52 & $-$101 $\pm$ 45 \\
136\tablenotemark{*} &  0:56:43.08 & $-$72:52:17.22 & $+$ 23 $\pm$ 48 \\
137 &  0:26:05.65 & $-$73:23:10.61 & $+$ 38 $\pm$ 28 \\
138 & 0:34:10.03 & $-$70:25:19.64 & $-$15 $\pm$ 47 \\
139 & 0:24:28.98 & $-$70:09:29.40 & $+$  41 $\pm$ 22
 \enddata
\tablenotetext{a}{ Missing source names are polarized sources identified by the SFIND source finding algorithm which do not correspond to real sources in Stokes $I$ image.}
\tablenotetext{*}{The source is behind the SMC.}
\end{deluxetable} 
\clearpage

\begin{landscape}
\begin{deluxetable}{lrrrrrr}
\tabletypesize{\footnotesize}
\tablecolumns{8} 
\tablewidth{0pc} 
\tablecaption{Radio Pulsars in the Small Magellanic Cloud} 
\tablehead{   
\colhead {Name} & \colhead{RA (J2000)}\tablenotemark{a} & \colhead{DEC (J2000)}\tablenotemark{a}   & \colhead{DM (pc cm$^{-3}$)\tablenotemark{a}} & \colhead{$\rm{DM_{foreground}}$ (pc cm$^{-3}$)\tablenotemark{b} } & \colhead{$\rm{DM_{SMC~pulsar}}$ (pc cm$^{-3}$) \tablenotemark{c}}& \colhead{RM(rad m$^{-2}$)\tablenotemark{a}}
}

\startdata 
PSR J0045-7042 & 00:45:25.69 & -70:45:07.1 & 70 $\pm$ 3 &40.75 & 29 &\nodata  \\
PSR J0045-7319 & 00:45:33.16 & -73:19:03.0 & 105.4 $\pm$ 0.7 &41.21& 64.2 &-14 $\pm$ 27 \\
PSR J0111-7131 & 01:11:28.77 & -71:31:46.8 & 76 $\pm$ 3 &42.65& 33 & \nodata  \\
PSR J0113-7200 & 01:13:11.09 & -72:20:32.2 & 125.49 $\pm$ 0.03 &42.92& 85.57 &\nodata \\
PSR J0131-7310 & 01:31:28.51 & -73:10:09.0 & 205.2 $\pm$ 0.7 &41.94& 163.3 &\nodata  \\
\enddata 
\label{table:radiopulsar}
\tablenotetext{a}{\cite{manchester2005,manchester2006}}
\tablenotetext{b}{NE 2001 Galactic Free Electron Model \citep{cordes2002}}
\tablenotetext{c}{Dispersion measure of pulsars after the removal of Galactic foreground contribution}
\end{deluxetable} 
\clearpage
\end{landscape}
\clearpage

\clearpage
\begin{deluxetable}{lr}
\centering
\tablecolumns{8} 
\tablewidth{0pc} 
\tablecaption{Rotation Measures of sources behind the SMC after foreground RM subtraction} 
\tablehead{   
\colhead{Source Name}   & \colhead{RM(rad m$^{-2}$)} }

\startdata 
018 &   $-$100 $\pm$  30 \\
019 & $-$60 $\pm$ 10 \\
020 &  $-$20  $\pm$  10 \\
021 &  $-$15 $\pm$  6 \\
025 &  $-$60 $\pm$  40 \\
121 & $-$90 $\pm$  40 \\
128 &  $-$230 $\pm$ 80 \\
134 & $-$400 $\pm$  60 \\
135 & $-$110 $\pm$  50 \\
136 &  0 $\pm$ 50  \\
\enddata 
\label{table:rmsources}
\end{deluxetable}

\begin{deluxetable}{lllcc}
\tablecolumns{5} 
\tablewidth{1000pt} 
\tablecaption{Foreground Corrected Starlight Polarization Data} 
\vspace{0.001in}
\tablehead{\colhead{Star ID\tablenotemark{a}} & \colhead{RA(hm)\tablenotemark{b}}   & \colhead{DEC(dm)\tablenotemark{b}}  & \colhead{Position Angle (Degrees)\tablenotemark{c}} & \colhead{Used in the C-F method?}}
\vspace{0.001in}
\startdata 
3 & 0:42.3 & $-$73:32 & $-$6 $\pm$ 90 & yes \\
7 & 0:44.9 & $-$73:48 & 41 $\pm$ 34 & yes \\
12 & 0:46.3 & $-$72:58 & 33 $\pm$ 25 & yes\\
18 & 0:47.0 & $-$73:17 & 48 $\pm$ 14 & yes\\
19 & 0:47.0 & $-$73:23 & 74 $\pm$ 13 & yes \\
23 & 0:47.5 & $-$73:12 & 40 $\pm$ 42& yes \\
26 & 0:48.0 & $-$72:9 & 9 $\pm$ 29 & yes\\
27 & 0:48.2 & $-$73:30 & 39 $\pm$ 62 & yes\\
31 & 0:49.0 & $-$73:4 & $-$10 $\pm$ 15 & yes\\
33 & 0:49.3 & $-$73:16 & $-$81 $\pm$ 6 & yes \\
35 & 0:49.4 & $-$72:46 & 18 $\pm$ 13 & yes\\
39 & 0:49.6 & $-$73:37 & $-$32 $\pm$ 31 & yes\\
40 & 0:49.7 & $-$73:36 & 17 $\pm$ 11 & yes\\
45 & 0:50.5 & $-$72:31 & 30 $\pm$ 9 & yes\\
46 & 0:50.5 & $-$72:15 & 19 $\pm$ 17 & yes\\
Anon(1) & 0:51.4 & $-$73:24 & 83 $\pm$ 10 & yes\\
54 & 0:52.0 & $-$73:15 & $-$4 $\pm$ 25 & yes \\
56 & 0:52.0 & $-$72:46 & $-$12 $\pm$ 19 & yes\\
55 & 0:52.2 & $-$71:47 & $-$11 $\pm$ 55 & yes\\
Anon(2) & 0:52.3 & $-$73:21 & 63 $\pm$ 16 & yes\\
59 & 0:54.0 & $-$72:17 & 18 $\pm$ 42 & yes\\
62 & 0:55.3 & $-$73:3 & 38 $\pm$ 14 & yes\\
68 & 0:56.7 & $-$71:28 & $-$61 $\pm$ 20 & yes\\
67 & 0:56.7 & $-$72:24 & 40 $\pm$ 41 & yes\\
74 & 0:57.9 & $-$72:34 & 17 $\pm$ 5 & yes\\
78 & 0:58.6 & $-$72:18 & 15 $\pm$ 27 & yes\\
80 & 0:58.6 & $-$72:19 & $-$19 $\pm$ 23 & yes\\
82 & 0:59.0 & $-$72:53 & 10 $\pm$ 124 & yes\\
85 & 0:59.3 & $-$72:22 & $-$35 $\pm$ 26 & yes\\
89 & 0:59.9 & $-$71:41 & 18 $\pm$ 149 & yes\\
97 & 1:0.5 & $-$72:41 & 83 $\pm$ 34 & yes\\
98 & 1:0.5 & $-$72:25 & 25 $\pm$ 92 & yes\\
103 & 1:1.2 & $-$72:44 & 20 $\pm$ 47 & yes\\
105 & 1:1.9 & $-$72:15 & $-$21$\pm$ 26 & yes\\
106 & 1:2.0 & $-$72:18 & $-$26 $\pm$ 28 & yes\\
107 & 1:2.1 & $-$71:56 & $-$20 $\pm$ 24 & yes\\
108 & 1:2.6 & $-$72:14 & $-$67 $\pm$ 26 & no\\
114 & 1:4.0 & $-$72:14 & $-$66 $\pm$ 131& no\\
117 & 1:4.1 & $-$72:16 & 20 $\pm$ 30 & yes\\
118 & 1:4.2 & $-$72:10 & 2 $\pm$ 13 & yes\\
119 & 1:4.2 & $-$72:48 & 33 $\pm$ 22 & yes\\
120 & 1:4.5 & $-$73:11 & 61 $\pm$ 36 & no\\
121 & 1:4.6 & $-$72:56 & $-$50 $\pm$ 44 & yes\\
124 & 1:5.2 & $-$72:27 & $-$57 $\pm$ 19 & yes\\
\newpage
128 & 1:5.7 & $-$72:30 & $-$37 $\pm$ 46 & yes\\
130 & 1:6.5 & $-$72:36 & $-$62 $\pm$ 12 & yes\\
136 & 1:8.2 & $-$73:11 & $-$7 $\pm$ 22  & yes\\
137 & 1:8.3 & $-$72:40 & $-$19 $\pm$ 20 & yes\\
138 & 1:8.5 & $-$72:33 & $-$86 $\pm$ 52 & no\\
145 & 1:10.7 & $-$72:39 & 20 $\pm$ 88 & yes\\
146 & 1:11.0 & $-$72:15 & $-$47 $\pm$ 28 & yes\\
149 & 1:11.8 & $-$71:17 & $-$6 $\pm$ 10 & no\\
150 & 1:12.3 & $-$72:54 & $-$12 $\pm$ 17 & yes\\
152 & 1:12.8 & $-$73:28 & $-$72 $\pm$ 21 & yes\\
154 & 1:13.6 & $-$73:21 & 21 $\pm$ 13 & yes\\
155 & 1:14.2 & $-$73:28 & 14 $\pm$ 11 & yes\\
156 & 1:14.8 & $-$73:28 & 84 $\pm$ 44 & no\\
157 & 1:15.2 & $-$73:29 & 40 $\pm$ 26 & yes\\
159 & 1:15.3 & $-$73:29 & 51 $\pm$ 55 & yes\\
163 & 1:19.2 & $-$72:48 & $-$13 $\pm$ 32 & yes\\
165 & 1:19.8 & $-$73:14 & 44 $\pm$ 98 & yes\\
168 & 1:21.1 & $-$72:54 & 77 $\pm$ 18 & yes\\
166 & 1:21.2 & $-$74:6 & $-$17 $\pm$ 26 & yes\\
181 & 1:28.7 & $-$72:50 & 20 $\pm$ 13 & yes\\
187 & 1:30.3 & $-$73:31 & $-$1 $\pm$ 23 & yes\\
189 & 1:30.7 & $-$72:56 & 46 $\pm$ 36 & yes\\
191 & 1:41.2 & $-$73:58 & $-$19 $\pm$ 26 & no\\
193 & 1:43.8 & $-$74:47 & 32 $\pm$ 7 & no\\
194 & 1:44.7 & $-$74:39 & 8 $\pm$ 28 & no\\
196 & 1:48.8 & $-$74:8 & $-$2 $\pm$ 21& no\\
198 & 1:50.8 & $-$74:3 & 69 $\pm$ 11 & no\\
202 & 1:52.8 & $-$74:2 & 50 $\pm$ 29 & no\\
205 & 1:53.2 & $-$74:10 & 18 $\pm$ 120 & no\\
215 & 2:13.3 & $-$74:38 & $-$19 $\pm$ 15 & no\\
HV 821 & 0:40.9 & $-$73:52 & 36 $\pm$ 23 & yes\\
NGC 330 & 0:55.5 & $-$72:37 & 25 $\pm$ 15 &  no \\*
\enddata 
\tablenotetext{a}{The star IDs have been taken from the catalogues of \cite{sanduleak1968a,sanduleak1968b,sanduleak1969} whenever available.}
\tablenotetext{b}{The positions of stars are in epoch 1975.}
\tablenotetext{c}{The position angle is measured counterclockwise with respect to the great circle joining the SMC to the LMC.}
\label{table:starlightdata}
\end{deluxetable}
\clearpage

\clearpage
\begin{deluxetable}{lcccc}
\tablecolumns{6} 
\tablewidth{0pt} 
\tablecaption{Measurement of the line of sight magnetic field strength through the SMC} 
\tablehead{   
\colhead{Source Name}  & \colhead{Model 1 $\overline{B_{\parallel}}$ ($\mu$G)}  & \colhead{Model 2 $\overline{B_{\parallel}}$ ($\mu$G)}  & \colhead{Model 3 $\overline{B_{\parallel}}$ ($\mu$G)}}
\startdata 
018 & $-$0.8$\pm$0.2      & $-$0.2$\pm$0.1      & $-$0.4$\pm$0.1\\
019 & $-$0.4$\pm$0.1      & $-$0.4$\pm$0.1      & $-$0.4$\pm$ 0.1\\
020 &  $-$0.1$\pm$0.1     & $-$0.1$\pm$0.1      & $-$0.1$\pm$0.1\\
021 &  $-$0.10$\pm$0.05 & $-$0.10$\pm$0.04 & $-$0.10$\pm$0.04\\
025 &  $-$0.4$\pm$0.3     & $-$0.1$\pm$0.1      &  $-$0.2$\pm$0.2\\
121 &   $-$0.7$\pm$0.3    & $-$0.2$\pm$ 0.1     & $-$0.4$\pm$0.2\\
128 & $-$1.8$\pm$0.6      & $-$0.5$\pm$0.2      &  $-$1.0$\pm$0.3\\
134 &  $-$3.0$\pm$0.4     & $-$0.09$\pm$0.01 & $-$0.5$\pm$0.1\\
135 &  $-$0.9$\pm$0.3     & $-$2.7$\pm$1.1      & $-$1.5$\pm$ 0.6\\
136 &  0.0$\pm$0.4           &      0.0$\pm$0.1          & 0.0$\pm$0.2
\enddata 
\label{table:b}
\end{deluxetable} 
\clearpage

\clearpage

\clearpage

\clearpage

\clearpage

\clearpage

\end{document}